\documentclass[11pt]{article}
\usepackage{comment}  
\usepackage{amsmath}
\usepackage{multirow}
\usepackage{amsthm}
\usepackage{amssymb}
\usepackage{graphics}
\usepackage{graphicx}
\usepackage{latexsym}
\usepackage{indentfirst}
\usepackage{graphicx}
\usepackage{amsmath}
\usepackage{amssymb}
\usepackage{amsthm}
\usepackage{bm}
\usepackage{relsize}
\usepackage{titling}
\usepackage{titlesec}
\usepackage{tikz}
\usepackage{float}
\usepackage{subcaption}
\usetikzlibrary{arrows}
\usepackage{hyperref}
\usepackage[export]{adjustbox}

\newtheorem{theorem}{Theorem}

\usepackage[final]{neurips_2019}

\usepackage[utf8]{inputenc} 
\usepackage[T1]{fontenc}    
\usepackage{hyperref}       
\usepackage{url}            
\usepackage{booktabs}       
\usepackage{amsfonts}       
\usepackage{nicefrac}       
\usepackage{microtype}      
\usepackage{graphicx}       
\usepackage{subcaption}
\usepackage{physics}        
\usepackage{color}
\usepackage{float}
\usepackage{comment}
\title{A self-supervised neural-analytic method to assess the evolution of COVID-19 in Romania}

\author{
  Radu D. Stochitoiu \\
  {\small Faculty of Automatic Control and Computers}\\
  {\small University Politehnica of Bucharest} \\
  {\small radu.stochitoiu@gmail.com} \\
   \And
   Marian Petrica \\
   {\small Gheorghe Mihoc - Caius Iacob Institute of}\\
   {\small Mathematical Statistics and Applied Mathematics}\\
   {\small of the Romanian Academy}\\
   {\small Faculty of Mathematics and Computer Science}\\
   {\small University of Bucharest}\\
   {\small marianpetrica11@gmail.com} \\
   \And
   Traian Rebedea \\
   {\small Faculty of Automatic Control and Computers}\\
   {\small University Politehnica of Bucharest}\\
   {\small traian.rebedea@cs.pub.ro} \\
   \And
   Ionel Popescu \\
   {\small Faculty of Mathematics and Computer Science}\\
   {\small University of Bucharest} \\
   {\small Institute of Mathematics of the} \\
   {\small Romanian Academy} \\
   {\small ionel.popescu@fmi.unibuc.ro} \\
   {\small ionel.popescu@imar.ro} \\
   \And
   Marius Leordeanu \\
   {\small Institute of Mathematics of the} \\
   {\small Romanian Academy} \\
   {\small University Politehnica of Bucharest} \\
   {\small leordeanu@gmail.com} \\
}

\begin{document}

\maketitle

\begin{abstract}
  Analyzing and understanding the transmission and evolution of an infectious disease is crucial to be able to design the best social and medical policies, foresee their results and deal with all subsequent social and economic effects. We address this important problem from a computational and machine learning perspective. More specifically, we want to statistically estimate all relevant parameters of the spread of COVID-19, such as the reproduction number, the death rate or the duration of the infectiousness period, based on Romanian patients, as well as to be able to predict future outcomes. This endeavor is important, since it is well known that these factors vary across the globe and might be dependent on many causes, including social, medical, age and genetic factors. At the core of our computational approach lies the paper, state-of-the-art work~\cite{chowdhury2020dynamic} that proposes an improved version of SEIR, which is one of the classic and established models for infectious diseases. We want to infer all the parameters of the model that govern the evolution of the pandemic based on the most reliable measurement, which is the number of deaths. The true number of infected people is impossible to precisely know.\\
  Once the model parameters are estimated, we are able to predict all the other relevant measures, such as the number of exposed and infectious people and many other factors, as shown in this paper. To achieve this, we propose a self-supervised approach to train a deep convolutional network to estimate the correct set of Modified-SEIR model parameters, given the observed number of daily fatalities. Lastly, starting from these initial parameters, we refine the solution with a stochastic coordinate descent approach. We compare our deep learning optimization scheme with the classic grid search approach and show great improvement in both computational time and prediction accuracy. We find an optimistic result in the case of fatality rate for Romania which may be around 0.3\% and we also demonstrate that our model is able to correctly predict the number of daily fatalities for up to three weeks in the future, while staying around the intervals defined by (\cite{Youyanggu}) that was also used in the United States of America and the predictions from IHME (\cite{2020.04.21.20074732}).
\end{abstract}

\section{Introduction}

Early understanding of the dynamics of an infectious disease is fundamental to being able to act in time and take the best safety measures for the population. Existing powerful mathematical models based on differential equations are able to assess reasonably well the evolution of the different curves (e.g. number of infected and hospitalized people, fatalities), given the correct set of model parameters. However, the inverse learning problem of finding the best parameters given the observed data is not an easy task, especially when the problem is not convex and several distinctive sets of parameters constitute relatively good local optima. The existence of multiple solutions is an important aspect, as we want to be able to predict future outcomes and learn fundamental parameters, such as the reproduction number and the fatality rate.

We take a dual neural-analytic approach, which effectively combines the power of the analytical solutions to model and predict data with a relatively small set of meaningful parameters with the power of deep neural networks to learn the inverse problem, that of estimating the correct parameters from the observed data. We propose an effective self-supervised training and prediction scheme, in which the two pathways, one classic, analytical (using differential equations) and the other based on machine learning (using deep convolutional networks) can feed each other, in tandem. During the self-supervised training phase, we start from random parameters of an improved, state-of-the-art modified SEIR model (\cite{chowdhury2020dynamic}, ~\cite{gabgoh}) (which we will refer to as Modified-SEIR), to predict fatality curves for a given period. Then we train the neural network to predict the known generator set of parameters, given the generated curves, using the analytical model. At test time, the network is used to rapidly estimate the correct set of parameters from the real, observed, curve of daily fatalities. Then, the set of parameters is further optimized by stochastic coordinate descent to minimize the L2-norm between the predicted curve (by the analytical Modified-SEIR model~\cite{chowdhury2020dynamic}) and the real curve of daily fatalities.  Note that we estimate the correct set of model parameters solely from the number of fatalities, since, as mentioned above, that number is the only one that could be measured correctly. The number of true infected people is impossible to know, given mainly the limitations in testing and the relatively large portion of the population that is asymptomatic.

Among the first measures taken by the authorities in Romania was to impose a very strict social distancing plan. This restriction has a major impact on the basic reproduction number of the SEIR model, generally reducing it by a percentage between 40\% and 80\% (\cite{socialDistancing}, \cite{Read2020.01.23.20018549}). As social distancing norms were alleviated on 15 May 2020, we considered two simulated scenarios, so we could analyze and compare the impact of the heavy vs. moderate isolation restrictions. We now know, based on \cite{when-to-lift}, that there are Gaussian simulations that could help us make a more educated guess for a date when to lift the lockdown completely.

The two main contributions that we make in this paper are:

\begin{enumerate}
    \item For prediction, we implement a proposed mathematical model based on SEIR (\cite{chowdhury2020dynamic, gabgoh}) (referred to as Modified-SEIR) to estimate the evolution of infectious diseases. We optimize the model to fit the data provided by Romanian health authorities for the COVID-19 pandemic, using a novel deep learning approach, trained in a self-supervised fashion.
    
    \item We introduce a novel self-supervised deep learning approach for fast optimization and learning of the parameters of the anlytical Modified-SEIR model. More specifically, the convolutional network is trained on many outputs of the Modified-SEIR model, generated by random sets of parameters, to predict precisely the same generator parameters set, in each case. After being trained on hundreds of thousands of such synthetic cases, the neural network becomes able to take us directly in the neighborhood of the best fitting parameters, when presented with the real data curve. Then, a refinement coordinate descent procedure is applied at the end, to obtain the final solution. Our experiments clearly show that the proposed deep learning approach to optimization greatly improves speed and accuracy over the baseline optimization approach (grid search with coordinate gradient).
\end{enumerate}

\section{Related work} \label{related-work}

Shortly after the first cases of COVID-19 appeared in the world, an important research movement began, which aims at finding bounds for the characteristics of the infectiousness of this new coronavirus. In \cite{Kucharski2020.01.31.20019901} authors show that the basic reproduction number is significantly influenced by travel restrictions, ranging from $2.35$ to $1.05$. In their procedures, they used an estimate of the incubation period equal to $5.2$ days, but which can be as low as $2$ days, according to a study conducted in Wuhan (\cite{doi:10.1056/NEJMoa2001316}). This study also found that the value of $2.2$ is a good approximation for the basic reproduction number, which is also similar to the findings of our research reported here.

In \cite{WU2020689} authors use a SEIR (Susceptible-Exposed-Infectious-Recovered) model to estimate the basic reproduction number at $2.68$ on January 25 in Wuhan. At the same time, it reveals a worldwide incubation time that ranges from $4.6$ to $6.9$. In this article we see the positive correlation between the basic reproduction number and the probability of creating an exponential epidemic starting with a single infected person.

The official report of the World Health Organization (\cite{who-report}) introduces several observed parameters, including a basic reproduction number in the range of $2-2.5$, an incubation period with an average in the range of $5-6$, a minimum hospitalization rate around $6.1 \%$, represented by critically ill patients, and a maximum of $\simeq 20 \%$, to which are added those $13.8 \%$ in severe condition, a period from incubation to death in a wide range of $2-8$ weeks and a recovery time for mild cases topped by $12$ days and for severe cases up to $14$ days of hospitalization.

In several studies dealing with the estimation of the incubation period of viruses such as 2019-nCoV, SARS or MERS (\cite{PMID:32046819}, \cite{article:lau}, \cite{article:virlogeux}) we notice values in the range of $4.4-6.9$.

Another important factor studied is how strict the rules of social distancing must be in order to stop the number of cases from increasing. In \cite{Read2020.01.23.20018549} it is estimated that a reduction in the basic reproduction number by $58-76 \%$ is needed to stop the increase in the number of infected people, considering the base reproduction number equal to $3.11$.

The core of our prediction model is the Modified-SEIR (\cite{chowdhury2020dynamic}), with a freely available toolbox, the Epidemic Calculator (\cite{gabgoh}), which we use as a stepping stone in our own implementation and validation of our model and experiments. 

To solve a very difficult prediction problem, such as the one relating to COVID-19, \cite{how-artificial} proposes AI-powered ways to manage limited healthcare resources, develop personalized patient management and treatment plans, inform policies, enable effective collaboration, and better understand and account for uncertainty.

A limited number of published medical studies~\cite{popescu2020covid, gherghel2020romania} present and discuss the epidemiology, clinical preparedness, and medical challenges of the COVID-19 pandemic in Romania. As shown next, our model accurately fits the real data, predicts future events (not seen during training) and estimates important core characteristics of the virus specific to Romania, such as the reproduction number, the length of infectiousness, time to recovery, and fatality rate. Note that the related work presented above provides specific ranges, established in the medical literature, within which we optimize the parameters of the Modified-SEIR.

Numerous studies have addressed the global evolution of the pandemics. In the context of the COVID-19 pandemic, the characterization and forecasting of transmission dynamics, as well as the rigorous estimation of model parameters, have been comprehensively examined in the papers \cite{Krivorotko, MASSONIS2021441, Ciupe, Long03082021, petrica2022regime, MARINOV2020100041, Sikder}.

\section{A Modified-SEIR prediction model}\label{model}

\begin{figure}[H]
  \centering
  \includegraphics[width=0.9\linewidth]{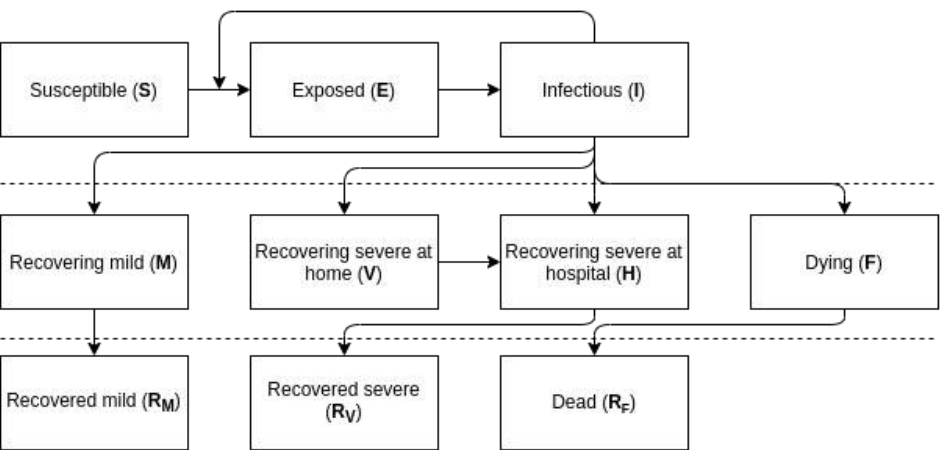}
  \caption{Modified SEIR model: the "Removed" case  is split into: 1) Recovered from mild symptoms; 2) Recovered from severe symptoms; 3) Deceased. The diagram follows the differential equations of the Modified-SEIR model presented in Table \ref{seir-eqs-table}.}
  \label{visual-seir}
\end{figure}

In order to assess and understand key factors of COVID-19 we approach the recent 
mathematical model (\cite{chowdhury2020dynamic}) based on the classic SEIR (\textbf{S}usceptible, \textbf{E}xposed, \textbf{I}nfectious, \textbf{R}emoved; \cite{doi:10.1137/S0036144500371907}), which is a widely accepted standard for modeling the evolution of infectious diseases. The Modified-SEIR model follows the usual steps in which an infectious disease evolves. The evolution along with key elements and measures are fully described by a set of differential equations (Table \ref{seir-eqs-table})
that we present in this section. We also offer a visual representation of the model in Figure~\ref{visual-seir}.

\newpage

We further divide the \textit{Removed} compartment into three categories: recovered from mild symptoms ($R_M$), recovered from severe symptoms ($R_V$) and deceased ($R_F$). There is also an extra layer of differential equations in the middle which helps us better shape and understand the dynamics of the disease. Following the Modified-SEIR model (\cite{chowdhury2020dynamic}) and the Epidemic Calculator (\cite{gabgoh}), we assume that all fatalities come from hospitals, and that all severe cases are admitted to hospitals immediately after the infectious period ends.

\begin{table}[H]
  \caption{Modified SEIR: System of differential equations, which describe the evolution of key characteristics and measures of pandemics over time. In our work we consider the number of fatalities ($F$, $R_F$) as the only real variable that could be measured correctly and we optimize the model according to it.}
  \label{seir-eqs-table}
  \centering
  \renewcommand{\arraystretch}{2}
  \begin{tabular}{ll}
    $\dv{S}{t} = - \beta I S$ & $\dv{E}{t} = \beta I S - \sigma E$ \\
    $\dv{I}{t} = \sigma E - \gamma I$ & $\dv{M}{t} = P_M \gamma I - \frac{1}{T_M} M$ \\
    $\dv{V}{t} = P_V \gamma I - \frac{1}{T_H} V$ & $\dv{H}{t} = \frac{1}{T_H} V - \frac{1}{T_V} H$ \\
    $\dv{F}{t} = P_F \gamma I - \frac{1}{T_F} F$ & $\dv{R_M}{t} = \frac{1}{T_M} M$ \\
    $\dv{R_V}{t} = \frac{1}{T_V} H$ & $\dv{R_F}{t} = \frac{1}{T_F} F$ \\
    $\beta = 
    \begin{cases}
        \frac{R_0}{T_{inf}}, & \text{before } T\\
        \frac{(1-P_T) R_0}{T_{inf}}, & \text{otherwise}
    \end{cases}$ & $\sigma = \frac{1}{T_{inc}}$ \\
    $\gamma = \frac{1}{T_{inf}}$ & $P_M = 1 - P_V - P_F$ \\
  \end{tabular}
\end{table}

The advantage of having an explicit analytic model versus a pure deep learning approach
is that each parameter has a clear meaning that is easy to interpret. 
In this case, the meaning of each parameter used is described and summarized in Table~\ref{seir-params-table}. The ranges in which we look for optimal parameters are fixed according to recently published medical research and their confidence intervals, as presented in Section \ref{related-work}. Note that we use constant values for some parameters, such as the total size of the population or the time from severe symptoms to hospitalization (according to \cite{gabgoh}).

We choose a continuous‐time ODE framework considering the following reasons:

\begin{itemize}
    \item ODEs admit well-known tools for steady‐state and stability analysis (eigenvalues, manifolds) that underpin our sensitivity study and control formulation.
    \item Embedding the continuous dynamics directly in the loss function allows back-propagation through time via adjoint or neural-ODE methods, which is more difficult in a discrete model.
    \item Although data arrive at daily intervals, the underlying processes (transmission, progression, recovery) occur continuously. A continuous model naturally averages over day-to-day variability and mitigates day-to-day reporting noise.
\end{itemize}

\begin{table}[H]
  \caption{Modified SEIR parameters. "Deduced" means that the values are deduced from the system of differential equations found at Table \ref{seir-eqs-table}.}
  \label{seir-params-table}
  \centering
  \begin{tabular}{llll}
    \toprule
    Name & Description & Initial value & Range \\
    \midrule
    $S$ & Susceptible population & $N - I_0$ & Deduced \\
    $E$ & Exposed population & $0$ & Deduced \\
    $I$ & Infectious population & $I_0$ & Deduced \\
    $M$ & Recovering at home with mild symptoms & 0 & Deduced \\
    $V$ & Recovering at home with severe symptoms & 0 & Deduced \\
    $H$ & Recovering in hospital with severe symptoms & 0 & Deduced \\
    $F$ & Dying & 0 & Deduced \\
    $R_M$ & Recovered from mild symptoms & 0 & Deduced \\
    $R_V$ & Recovered from severe symptoms & 0 & Deduced \\
    $R_F$ & Dead (Fatal) & 0 & Deduced \\
    $P_M$ & Mild symptoms rate & & Deduced \\
    $P_V$ & Severe symptoms rate & & $[0.04 - 0.2]$\\
    $P_F$ & Case fatality rate & & $[0.1\% - 3\%]$\\
    $T_{inc}$ & Length of incubation period (days) & & $[2-14]$\\
    $T_{inf}$ & Length of infectiousness period (days)  & & $[3-14]$\\
    $T_M$ & Recovery time for mild cases (days) & & $[4-12]$\\
    $T_V$ & Recovery time for severe cases (days) & & $[7-35]$\\
    $T_H$ & Time from severe symptoms onset to hospitalization (days) & & $[5]$\\
    $T_F$ & Time from end of infectiousness to death (days) & & $[14-35]$\\
    $R_0$ & Basic reproduction number & & $[1.5-2.5]$\\
    $T$ & Intervention time to reduce $R_0$ (days) & & $[20-22]$\\
    $P_T$ & Percentage to decrease transmission by after intervention & & $[40\% - 80\%]$\\
    $\beta$ & Transmission rate & & Deduced \\
    $\sigma$ & Rate of getting infectious from being exposed & & Deduced \\
    $\gamma$ & Recovery rate & & Deduced \\
    $I_0$ & Number of initial infections & & $[1500-2500]$ \\
    $N$ & Total size of population & & $[20175912]$\\
    \bottomrule
  \end{tabular}
\end{table}

In order to solve the time-based differential equations and produce the different evolution curves
we use a fourth order Runge-Kutta integrator (\cite{Tan2012}).

\section{Learning the model parameters} 
\label{sec:error-minim}

As the number of new fatalities per day is easily known and is not influenced by the number of actual tests run in any location, we consider it as ground truth in our experiments. Even though this number mixes patients with comorbidities 
(whose health is also influenced by other conditions) with those without comorbidities, the actual measured number of infected people who are dying is ultimately the only estimation that can be considered certain in the current research. We use the data uploaded daily by \href{https://github.com/CSSEGISandData/COVID-19/tree/master/csse_covid_19_data/csse_covid_19_daily_reports}{Johns Hopkins CSSE}.

Our purpose is to find a set of parameters that best approximates a curve modeled by the modified SEIR, the real curve of the daily number of fatalities. We denote $D_{real}$ as the vector (curve) of the reported daily number of deaths in a specific time interval and $D_{\theta}$ as the vector (curve) generated by the model with parameters $\theta$, for the same time interval. Thus, the cost function that we minimize to find the best fitting parameters is the square root of the sum of squared errors between the real and the predicted curve, as shown in Equation \eqref{eq:1}. We denote the optimal set of parameters by $\theta^*$, as defined in \eqref{eq:2}. The valid search (optimization) ranges for each parameter are presented in Table \ref{seir-params-table}, as mentioned previously, and constitute relatively large search regions, as unions over the ranges published in recent medical literature.

We test and compare two main ways to find the optimal parameters. One is a baseline, which starts with a classic grid search procedure followed by a stochastic coordinate descent refinement. The second optimization method, which is our main technical novelty, is to make an initial guess of the parameters using the self-supervised trained convolutional network followed by the same refinement step using stochastic coordinate descent. Each optimization module (grid search, neural network, and coordinate descent) is described in the next sections.

\begin{align}
    J(\theta) = \sqrt{\sum_i{(D_{real}(i) - D_{\theta}(i))^2}} \label{eq:1} \\
    \theta^* = \arg\min_{\theta} J(\theta) \label{eq:2}
\end{align}

Throughout calibration we use the first difference of the cumulative fatal compartment, $\,D(i)=\Delta R_F(i)=R_F(i)-R_F(i-1)$. The cost function in eq. \ref{eq:1} then measures the fit between this model-generated daily fatality curve and the reported data.

 All compartmental variables in our ODE system (e.g.\ $S(t),E(t),I(t),R_F(t)$) are defined as \emph{fractions} of the total population $N$.  To compare the model to observed daily‐death counts $D_{\text{real}}(i)$, we simply multiply the model’s fatality compartment by $N$:

    $$
      \widetilde{R}_F(t)\;=\;N\cdot R_F(t)
      \quad\Longrightarrow\quad
      D_\theta(i)\;=\;\Delta\widetilde{R}_F(i)
      \;=\;N\bigl[R_F(i)-R_F(i-1)\bigr].
    $$

\subsection{Optimization by grid search}
\label{sec:grid_search}

There are 11 parameters that we optimize over: $I_0$, $R_0$, $T_{inc}$, $T_{inf}$, $P_F$, $T_F$, $T_M$, $T_V$, $P_V$, $P_T$, $T$. Therefore, a full and very fine grid search is computationally infeasible. However, we can divide each range into (2-4) smaller ranges and look for a decent approximation to start with. The grid search module is followed by the coordinate descent refinement procedure (Section \ref{coordinate-descent}).

\subsection{Optimization by a neural network}
\label{sec:neural_net_optimization}

Because of the long computational time required by grid search optimization, we propose a deep learning approach, using a convolutional neural network trained in a self-supervised manner, as discussed previously, that is able to bring the solution to the neighborhood of the optimum very fast. The neural network optimization module is also followed by the same final coordinate descent procedure. Interestingly enough, it turns out that the results, when using the neural net optimization, are vastly superior in both speed and accuracy to the grid search approach.

\subsection{Self-supervised learning of neural net optimization}

Below we present the exact steps taken for the self-supervised scheme in which the neural network learns to guess the right set of parameters, given a curve of daily deaths for a given period of time.

\begin{enumerate}
    \item Create a dataset
    \begin{itemize}
        \item Take 100000 random samples from a uniform distribution of the 11 parameters in the ranges indicated in Table \ref{seir-params-table};

        \item Generate a curve of daily fatalities using the Modified-SEIR model (Table \ref{seir-eqs-table}) for each set of model parameters picked at the previous step. Although the estimated total number of deaths is cumulative (the $R_F$ variable), we take the number of daily deaths as daily increments in the total number of fatalities.
        
        \item For each curve, we randomly select a fixed number $L$ of consecutive days of daily deaths ($L$ is defined by the number of days in certain time ranges in our experiments, such that for [March 22, May 3], $L = 43$). This vector of $L$ consecutive numbers, representing the fatalities for the corresponding $L$ days, modeled by Modified-SEIR, along with the corresponding set of parameters, will constitute the 100K training pairs used in training the neural net optimizer presented in the following. Note that the start of the $L$-day sequence is chosen randomly (so it could be at the beginning or towards the end of the pandemic). We try to mirror the real case, when we really do not know which day should be considered the first of the pandemic.
    \end{itemize}
    
    \item Multi-head neural network optimizer, trained in a self-supervised way, to predict the Modified-SEIR model parameters, given the $L$-element curve of daily fatalities (generated by precisely the same set of parameters that should be predicted by the network).

    \begin{itemize}
        \item Hidden layers 
        \begin{enumerate}
            \item Conv1D (512, 5, 'relu')
            \item MaxPooling1D
            \item Conv1D (128, 5, 'relu')
            \item MaxPooling1D
            \item Conv1D (32, 5, 'relu')
            \item MaxPooling1D
            \item Flatten
            \item Dense (512)
            \item Dense (256)
            \item Dense (128)
        \end{enumerate}
        \item One output for each parameter
        \item Loss: Mean Squared Error
        \item Optimizer: Adam (\cite{adam-opti})
    \end{itemize}
\end{enumerate}

\begin{figure}[H]
  \centering
  \includegraphics[width=0.8\linewidth]{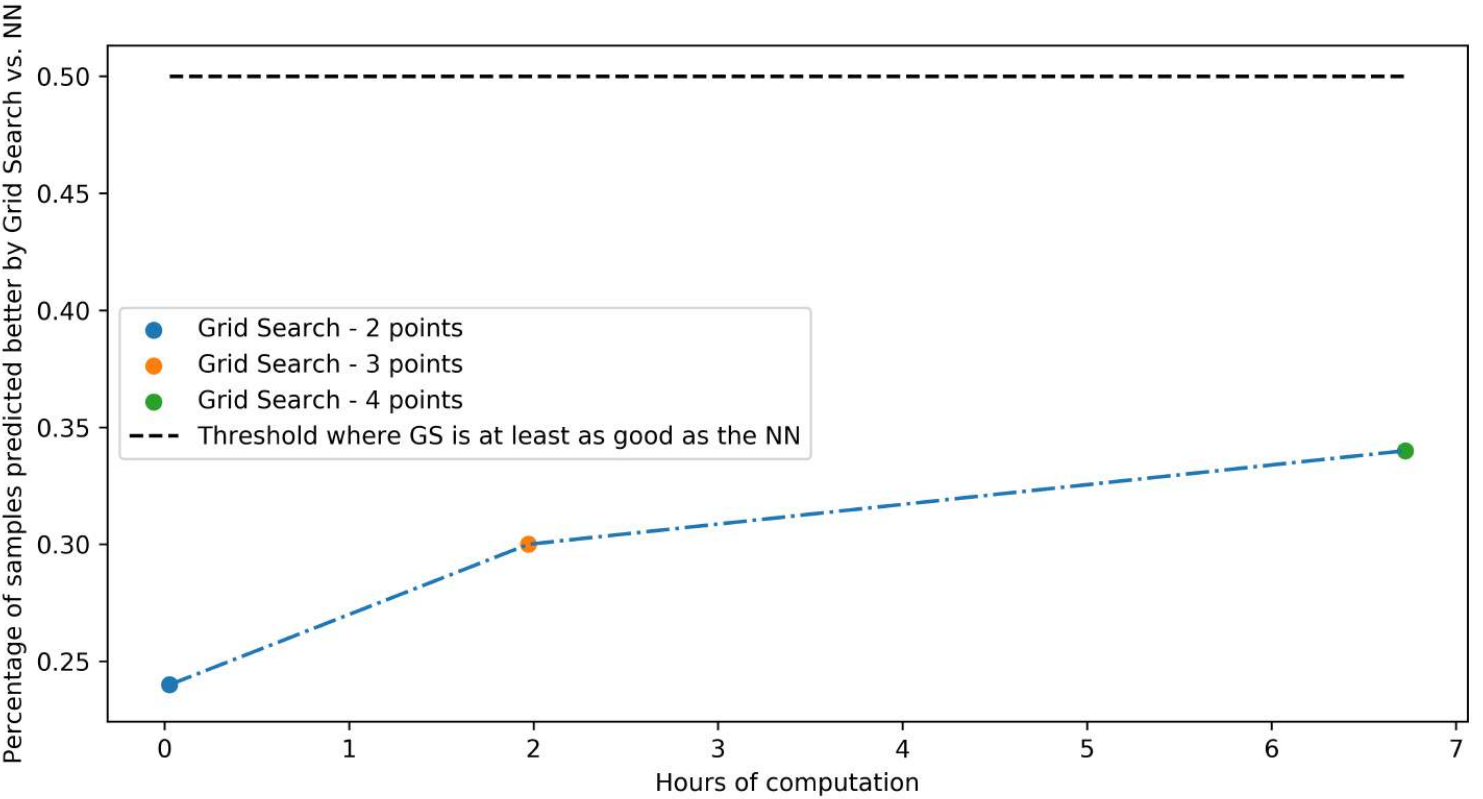}
  \caption{Comparison between the grid search and the neural network solutions. We present how often grid search produces better solutions than the neural net. We clearly see that 7 hours of computation are far from sufficient for grid search to beat the neural net. Note that in the plot, the final refinement procedure is not used by either of the two approaches.}
  \label{gs-nn-comparison}
\end{figure}

We compare the grid search approach with neural network predictions. The advantage of the latter approach is that it offers almost instantaneous predictions. In Figure \ref{gs-nn-comparison} we show the percentage of tasks (problems that are randomly generated by the model using random initialization) where the samples (sets of parameters) found by grid search produce superior curves (closer to the truth) than the ones predicted by the neural network. The percentage is a function of time since for an infinite amount of time we expect grid search, with a sufficiently fine grid, to beat the neural network. The plot shows how vastly superior in terms of speed the neural network is. Almost 7 hours of running for grid search, on an Intel© Core™ i9-9980HK CPU @ 2.40GHz x 8, are not enough to surpass the neural net predictions.

\subsection{Final coordinate descent refinement} 
\label{coordinate-descent}

We can expect that neither of the two approaches (grid search and neural net optimization) will directly produce an optimal solution, even though we do expect them to output a set of parameters that are close to a local optimum. In order to refine our results, we further apply an iterative \textit{stochastic coordinate descent} approach similar to \cite{doi:10.1007/s10107-015-0892-3}. Starting from the best predictions of a given first-stage module (neural net or grid search), we take random subsets of two parameters at a time, divide their search ranges into 20-40 parts around the current best solution,
and replace the next values of the chosen parameters with the ones that minimize the cost function. We iterate the procedure until we reach a convergence of $10^{-3}$ absolute error.

We make use of the stochastic property iteratively because we assume that some parameters influence the cost more than others. Thus, by choosing random subsets of parameters to optimize over, we avoid the risk of spending valuable time optimizing over subsets of parameters that do not bring much value.

While our model is formulated as a deterministic ODE system, one could extend it to an SDE. Such stochastic formulations enable explicit representation of reporting noise, but at the cost of significantly greater computational and inferential complexity. In particular, likelihood‐based calibration of SDEs often requires particle‐filtering or Monte Carlo methods, and analytical steady‐state and sensitivity‐analysis results become less tractable. We view the deterministic approach as a pragmatic compromise: it allows efficient gradient‐based training and clear stability insights, while still capturing the dominant epidemic trends observed in aggregate fatality data.

\section{Experimental analysis}

\subsection{On data dependency}

We consider three particular data sets on which we optimize and search for the best parameters of our model.

\begin{itemize}
    \item Daily fatalities from March 22 to May 3  (2020);
    \item Daily fatalities from March 22 to May 14 (2020);
    \item Daily fatalities from March 22 to May 21 (2020).
\end{itemize}

We know that on May 15 2020 the Romanian authorities changed the policy from state of emergency to state of alert. Because the time from incubation to death is greater than one week, we assume the reported daily fatalities from May 15 2020 to May 21 2020 are not influenced by the changed policy. This enables us to search for optimal parameters for the interval March 22 2020 to May 21 2020, as it can be modelled by Modified-SEIR.

For every data set, we present in Table \ref{table:data-dependency-params} the optimal model parameters found with neural net optimization followed by the refinement step, as defined in Section \ref{sec:error-minim}. As stated previously, they minimize the L2 distance between the real and the predicted curves of fatalities from March 22 to May 3, May 14, and May 21, respectively (all in the year 2020, of course). For each set of optimal parameters, we compute the error of prediction for the following dates: May 3, May 15, May 21, June 3, June 8, June 9, June 10, June 11 (all in 2020).

As the prediction errors are inversely proportional to the data set size, we prove that our approach is data-driven and provides better solutions for bigger data sets. In other words, every new day of observations is important in finding the real parameters that shape the evolution of the COVID-19 infectious disease.

\newpage

\begin{table}[H]
  \caption{Best parameters found by our neural net optimization followed by the final coordinate descent refinement.}
  \label{table:data-dependency-params}
  \centering
  \begin{tabular}{lllll}
    \toprule
    Name & Description & May 3 & May 14 & May 21 \\
    \midrule
    $I_0$ & Initial infectious population & $2450$ & $1600$ & $1600$\\
    $R_0$ & Basic reproduction number & $2.63$ & $2.63$ & $2$\\
    $T_{inc}$ & Length of incubation period (days) & $2$ & $2$ & $2$\\
    $T_{inf}$ & Length of infectiousness period (days) & $7.4$ & $7.4$ & $3$\\
    $P_F$ & Case fatality rate & $0.39\%$ & $0.68\%$ & $0.39\%$\\
    $T_F$ & Time from end of infectiousness to death (days) & $14$ & $16.1$ & $28$\\
    $T_M$ & Recovery time for mild cases (days) & $4$ & $13.28$ & $4$\\
    $T_V$ & Recovery time for severe cases (days) & $7$ & $7$ & $7$\\
    $P_V$ & Severe symptoms rate & $5\%$ & $15.6\%$ & $10\%$\\
    $P_T$ & Decrease in transmission after intervention & $60\%$ & $62\%$ & $56\%$\\
    $T$ & Intervention time to reduce $R_0$ (days)  & $21$ & $21$ & $21$\\
    \midrule
    Err. May 3 & Prediction absolute error on May 3, 2020 & $3.04\%$ & $3.92\%$ & $3.16\%$ \\
    Err. May 15 & Prediction absolute error on May 15, 2020 & $4.21\%$ & $3.55\%$ & $0.19\%$ \\
    Err. May 21 & Prediction absolute error on May 21, 2020 & $9.95\%$ & $8.22\%$ & $1.99\%$ \\
    Err. Jun 3 & Prediction absolute error on June 3, 2020 & $24.92\%$ & $20.14\%$ & $6.87\%$ \\
    Err. Jun 8 & Prediction absolute error on June 8, 2020 & $31.22\%$ & $24.94\%$ & $8.44\%$ \\
    Err. Jun 9 & Prediction absolute error on June 9, 2020 & $31.83\%$ & $25.26\%$ & $8.12\%$ \\
    Err. Jun 10 & Prediction absolute error on June 10, 2020 & $33.31\%$ & $26.40\%$ & $8.60\%$ \\
    Err. Jun 11 & Prediction absolute error on June 11, 2020 & $34.48\%$ & $27.25\%$ & $8.77\%$ \\
    \bottomrule
  \end{tabular}
\end{table}

\subsection{On finding optimal parameters} \label{section:optimality}

It is worth mentioning that in general the reversed problem of finding the correct model parameters from the observation of partial data (in our case we observe only the values of one output variable, the number of daily fatalities) is ill posed, since the problem is not necessarily convex and many different sets of parameters can produce similar output. However, this is a common case in machine learning, in which several valid solutions exist. The AI system usually learns to predict the most probable output (in this case, the set of parameters) given the observed data, based on its training experience. One good example is the case of vision, in which many different 3D worlds with different semantic interpretations could produce the same 2D image. Nevertheless, the visual system learns to pick the most likely interpretation (given its prior experience) out of infinitely many. In our specific case, we expect that the synthetic generation of many curves from different sets of parameters will help the neural network implicitly learn the priors in the data model, such that the network will learn to output, from the many different solutions, one that is most likely to have produced the given curve. We can easily imagine that there are certain distinct neighborhoods of parameters that generate similar curves and that a set of parameters coming from a larger neighborhood is more likely than one coming from a smaller one. In other words, sets of model parameters for which the output curve is more stable (low curve gradient w.r.t parameters) are probably more likely than sets of parameters for which the curve changes rapidly in their immediate neighborhood. Such subtle priors in the space of parameters should be implicitly learned during the self-supervised training if sufficiently many pairs of model-generated curves (input) - parameters (output) are presented to the network.

\begin{figure}[H]
    \centering
    \begin{subfigure}{0.49\textwidth}
        \centering
        \includegraphics[width=\linewidth]{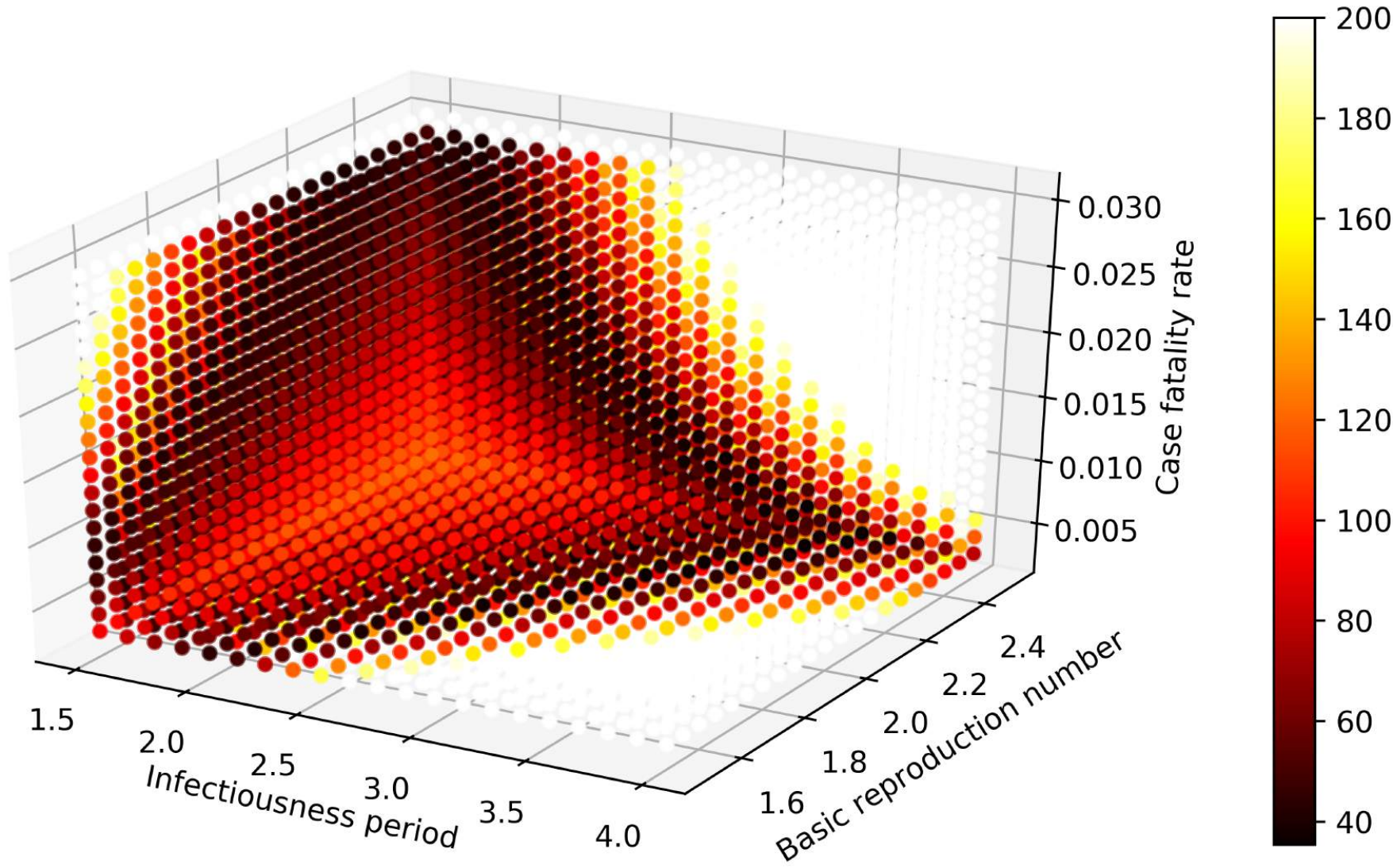}
    \end{subfigure}
    \hfill
    \begin{subfigure}{0.49\textwidth}
        \centering
        \includegraphics[width=\linewidth]{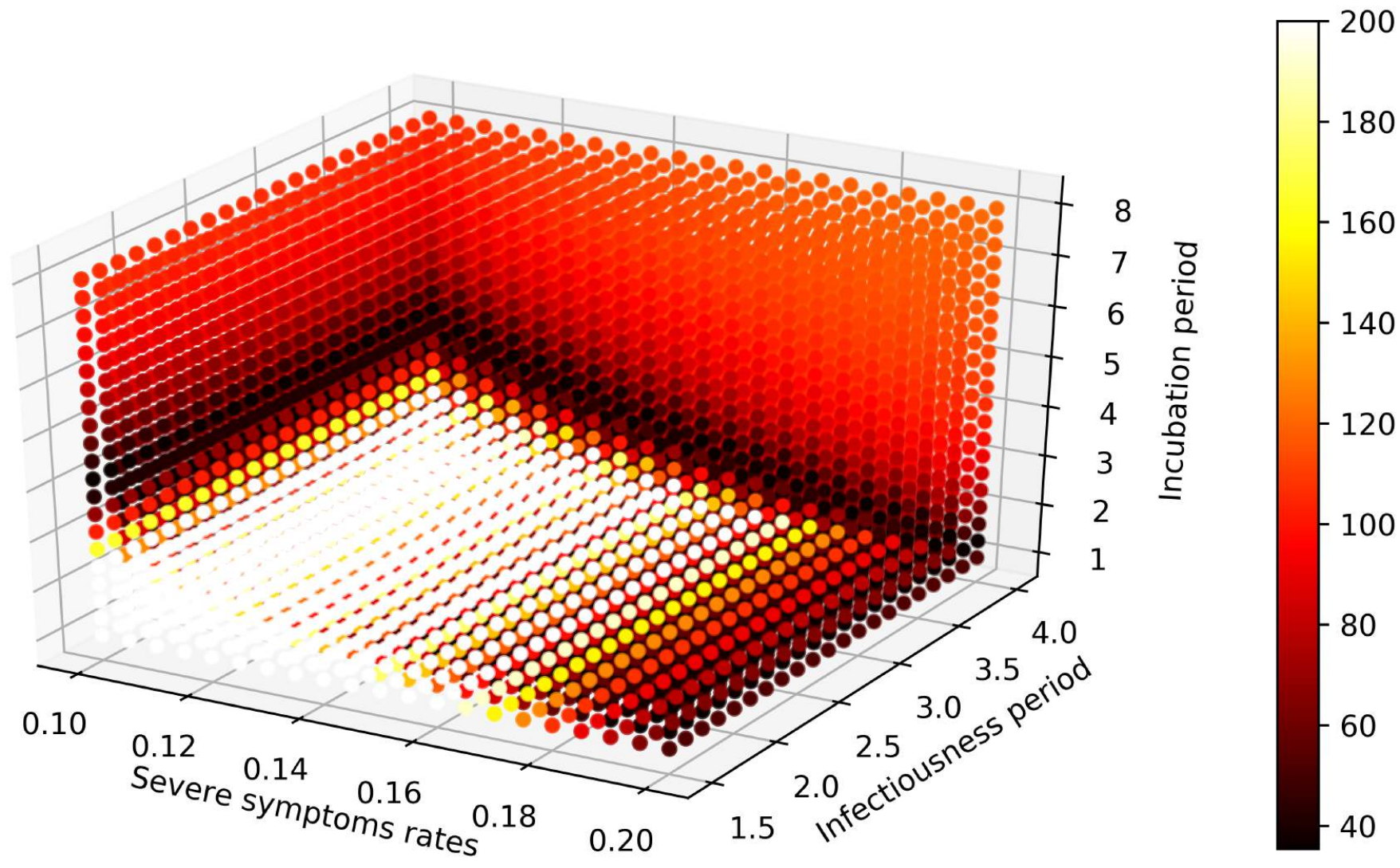}
    \end{subfigure}
    
    \vspace{0.5cm} 
    
    \begin{subfigure}{0.49\textwidth}
        \centering
        \includegraphics[width=\linewidth]{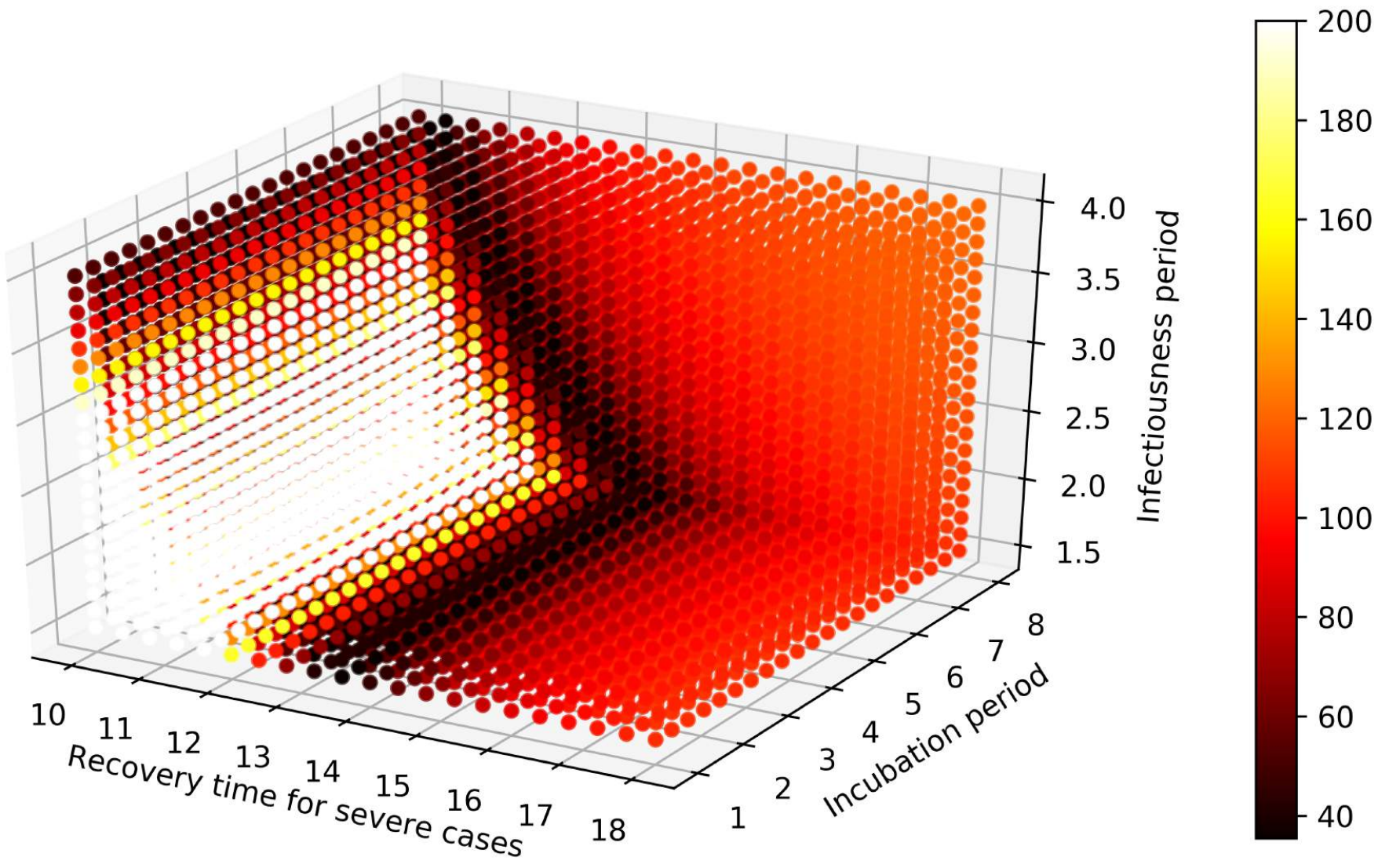}
    \end{subfigure}
    \hfill
    \begin{subfigure}{0.49\textwidth}
        \centering
        \includegraphics[width=\linewidth]{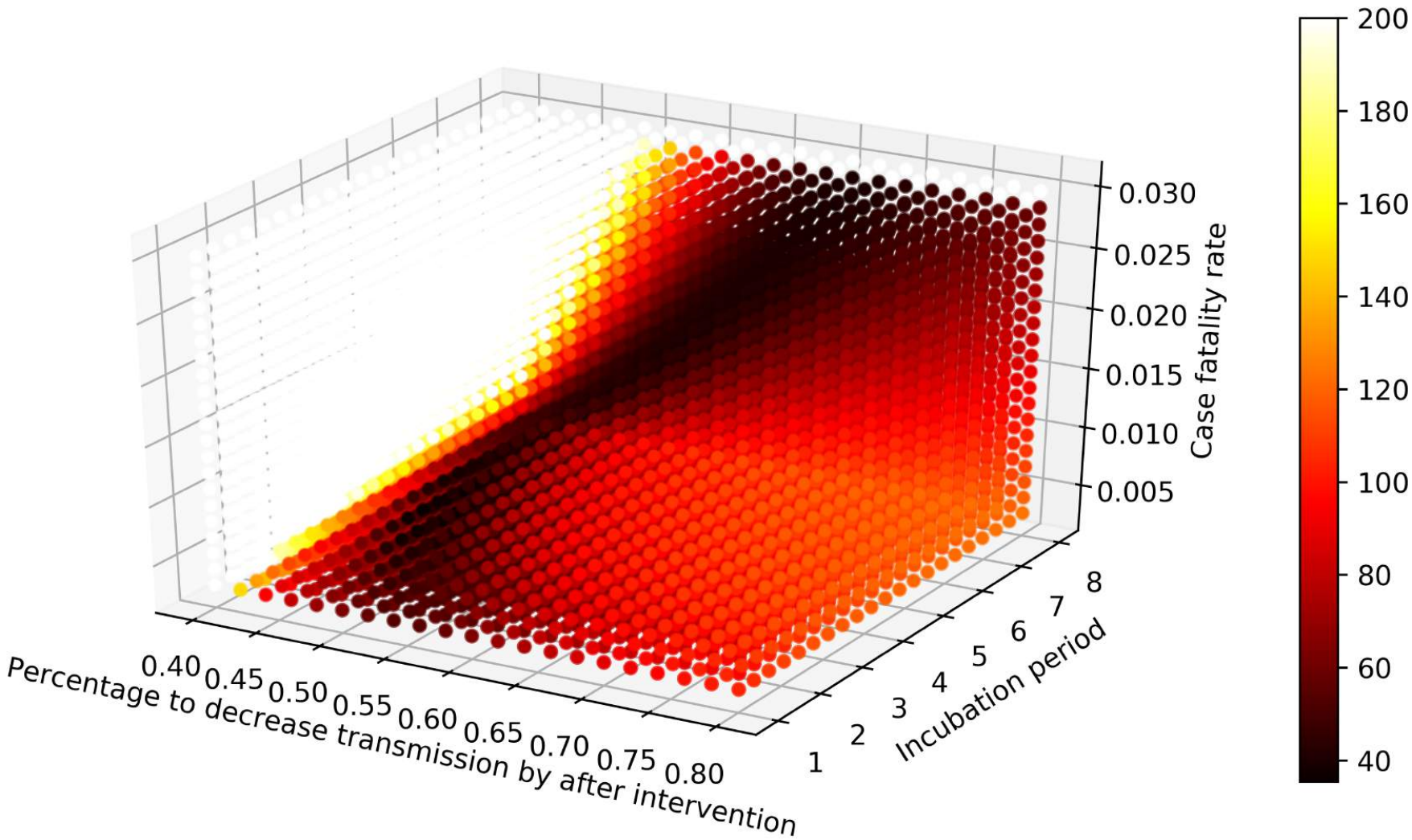}
    \end{subfigure}
    
    \caption{Plots of the cost function, represented by color, computed for joint distributions of three parameters at a time. The cost function goes from black (optimal/smallest) to white (least optimal/greatest). For better visualization we upper bound the cost at 200.}
    \label{fig:optimality}
\end{figure}

It is clear by now that searching for the best parameters is a complex task, especially if the cost function is not convex, thus admitting multiple local minima. Besides the intuitive discussion, we also experimentally analyze this issue by comparing different 4D plots where the 3D space is defined by three different parameters and the fourth dimension, the final cost function, is defined by color. In Figure \ref{fig:optimality} one can see that there are regions where there is a linear dependency between two parameters preserving the minimum cost and that there are multiple intervals with local minima for sets of three parameters. The latter finding is especially interesting as it tells us that our prediction might not be the correct one even if it has the lowest cost. Or, in other words, different distinctive sets of parameters may generate the same final cost. Thus, learning from larger sets of reported data, over longer periods of time, may shift the balance towards a different set of optimal parameters which can immediately modify the foreseen dynamics of the coronavirus infectiousness and fatality. This situation, which is commonly encountered in many AI tasks, reveals the inner ambiguity and difficulty of the problem tackled. What is, however, important and relevant here, is that we are able to learn sets of parameters which are plausible, often matching many independent findings in the literature, which predict with surprising accuracy the evolution of COVID-19 in Romania.

\begin{figure}[H]
  \centering
  \includegraphics[width=0.8\linewidth]{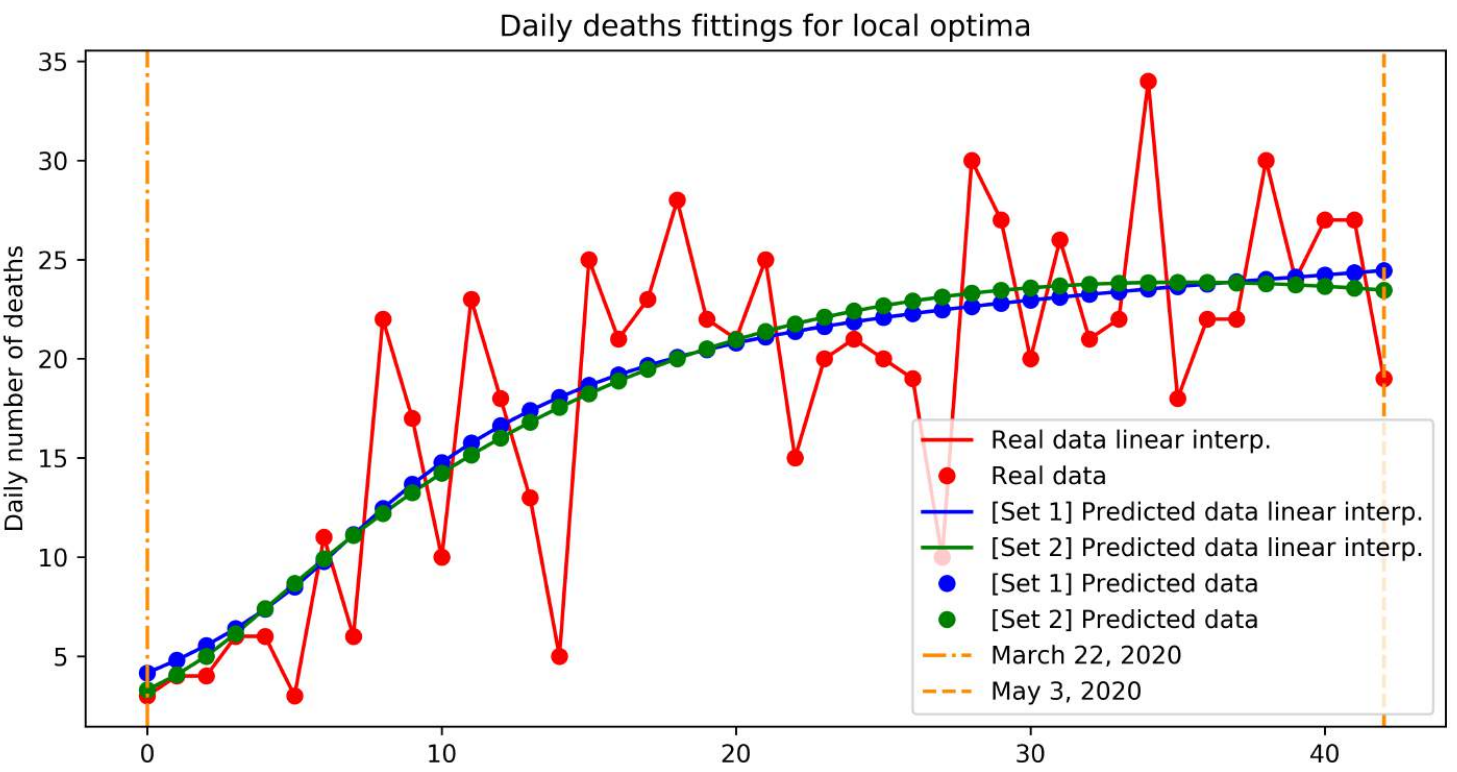}
  \caption{Daily deaths fitted curves for two sets of parameters with similar cost. The red dots represent the officially reported daily number of deaths in Romania. The blue and green lines with dots show the fitting of the Modified-SEIR models through the real data.}
  \label{fig:compare-daily-deaths}
\end{figure}

Starting from the knowledge that a particular parameter set might not necessarily be the correct one even if it has the lowest cost on the limited training data, we compare the predictions of two different sets of parameters that have similar $L^2$ costs 
(close to the lowest cost found for the interval from March 22, 2020, to May 3, 2020). In Figure \ref{fig:compare-daily-deaths} we present the evolution curves corresponding to these two sets of parameters. Both of them fit the given data equally well.

\newpage

\begin{figure}[H]
  \centering
  \includegraphics[width=0.8\linewidth]{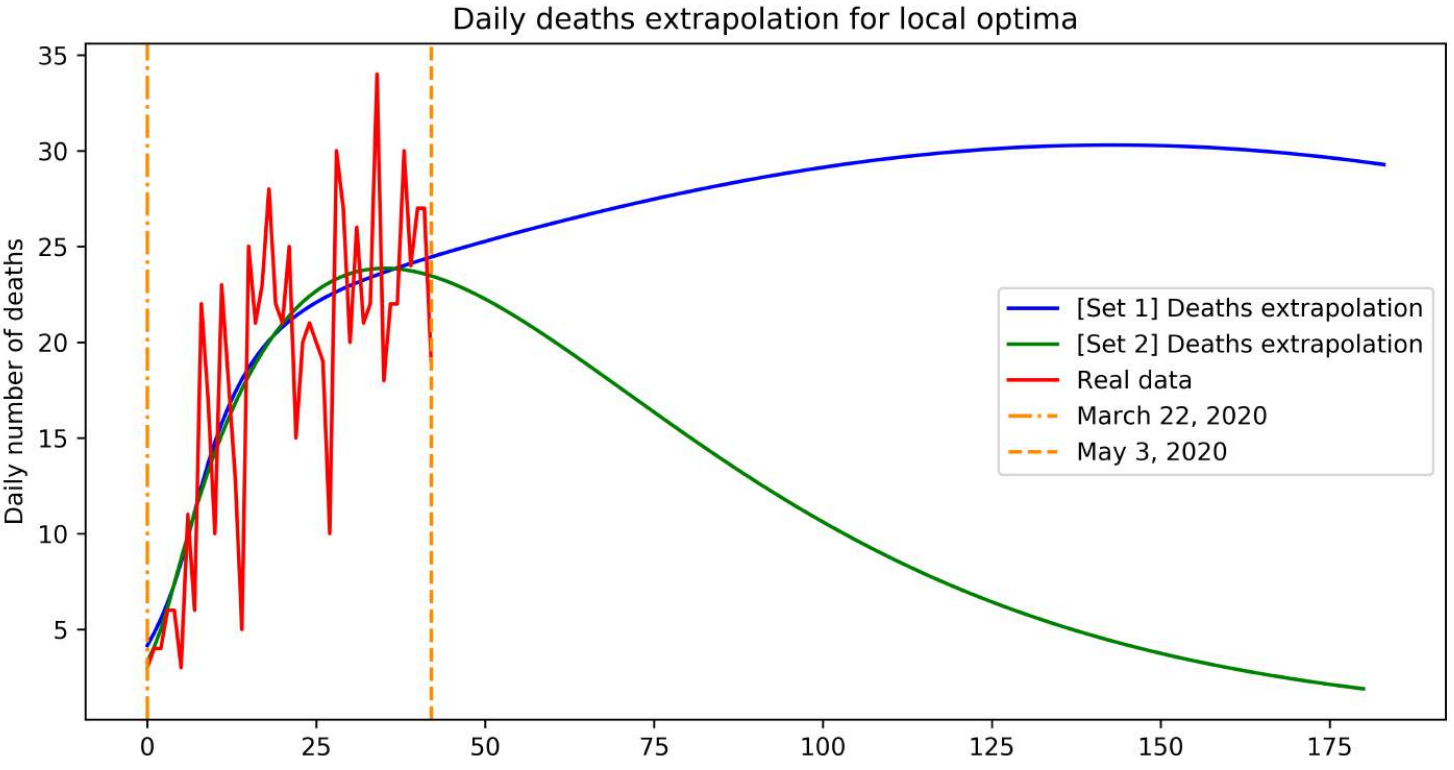}
  \caption{Daily deaths fitted curves extrapolation for two sets of parameters with similar costs (on seen data). The red dots represent the officially reported daily number of deaths in Romania. The blue and green lines show the extrapolation of the Modified-SEIR models through the real data. Note how different the future predictions are, between the two sets of parameters.}
  \label{fig:compare-daily-deaths-extrapolation}
\end{figure}

Surprisingly, when we extrapolate the curves into the future (by running the models according to the corresponding parameters into the future), we notice substantially different evolution curves. In Figure \ref{fig:compare-daily-deaths-extrapolation} we notice how one approximation goes down quickly, while the other continues to increase for another three months.

\begin{figure}[H]
  \centering
  \includegraphics[width=0.8\linewidth]{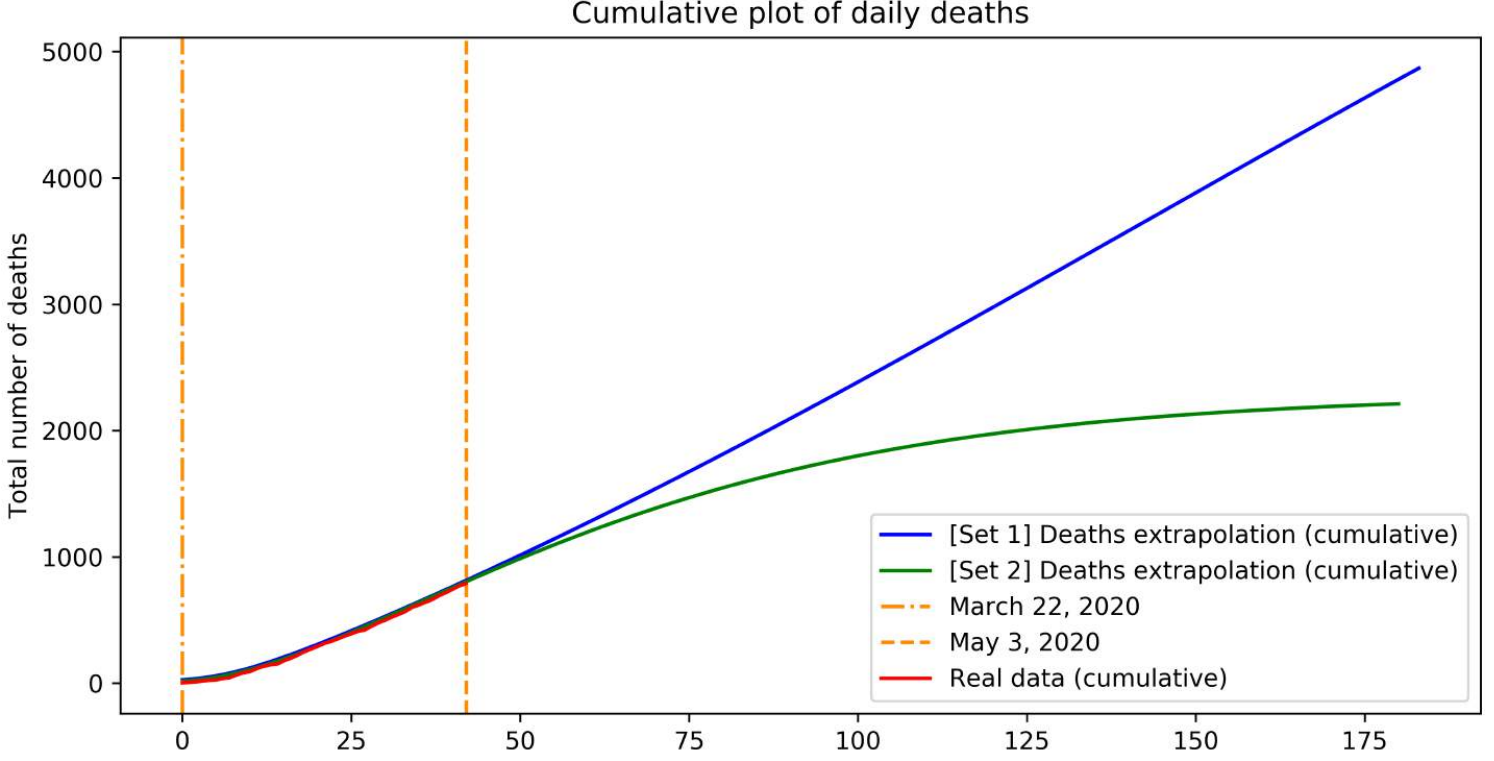}
  \caption{Cumulative extrapolation of the total number of fatalities for two sets of parameters with similar cost. The red dots represent the officially reported total number of deaths in Romania. The blue and green lines show the cumulative extrapolation of the Modified-SEIR models through the real data.}
  \label{fig:compare-cumulative-deaths}
\end{figure}

This subtle change in the way the two fitted curves diverge at the end of the data set has drastic outcomes regarding, among other characteristics, the number of active infections and the total number of fatalities. Figure \ref{fig:compare-cumulative-deaths} shows us that while the fitted curves for the observed interval (March 22, 2020, to May 3, 2020) are similar, one set of parameters predicts a total of 2290 deaths, while the other predicts a total of 10225. We compare their predictions in Table \ref{table:comparison}.

\begin{table}[H]
  \caption{Two sets of parameters for the Modified-SEIR model with similar costs but very different evolutions.}
  \label{table:comparison}
  \centering
  \begin{tabular}{llll}
    \toprule
    Name & Description & Set 1 & Set 2 \\
    \midrule
    $I_0$ & Initial infectious population & $2450$ & $1548$\\
    $R_0$ & Basic reproduction number & $2.63$ & $2.32$\\
    $T_{inc}$ & Length of incubation period (days) & $2$ & $2$\\
    $T_{inf}$ & Length of infectiousness period (days) & $7.4$ & $3.48$\\
    $P_F$ & Case fatality rate & $0.39\%$ & $0.15\%$\\
    $T_F$ & Time from end of infectiousness to death (days) & $14$ & $25.9$\\
    $T_M$ & Recovery time for mild cases (days) & $4$ & $10$\\
    $T_V$ & Recovery time for severe cases (days) & $7$ & $10$\\
    $P_V$ & Severe symptoms rate & $5\%$ & $10\%$\\
    $P_T$ & Decrease in transmission after intervention & $60\%$ & $59.2\%$\\
    $T$ & Intervention time to reduce $R_0$ (days) & $21$ & $21$\\
    \midrule
    $J(\theta)$ & L2 cost of the fitting & $33.285$ & $33.597$ \\
    Err. May 3 & Prediction absolute error on May 3, 2020 & $3.04\%$ & $1.65\%$\\
    Err. May 15 & Prediction absolute error on May 15, 2020 & $4.21\%$ & $0.28\%$\\
    Err. May 21 & Prediction absolute error on May 21, 2020 & $9.95\%$ & $5.99\%$\\
    Err. Jun 3 & Prediction absolute error on June 3, 2020 & $24.92\%$ & $10.80\%$\\
    Err. Jun 8 & Prediction absolute error on June 8, 2020 & $31.22\%$ & $13.22\%$\\
    Err. Jun 9 & Prediction absolute error on June 9, 2020 & $31.83\%$ & $13.07\%$\\
    Err. Jun 10 & Prediction absolute error on June 10, 2020 & $33.31\%$ & $13.75\%$\\
    Err. Jun 11 & Prediction absolute error on June 11, 2020 & $34.48\%$ & $14.02\%$\\
    \textbf{$\sum$ Fatalities} & \textbf{Prediction of total number of fatalities} & \textbf{10225} & \textbf{2065}\\
    \bottomrule
  \end{tabular}
\end{table}

\subsection{Key observations about the model parameters found}

As discussed in the previous section, the approach based on a neural network optimization followed by a stochastic alternative coordinate descent seems to overfit the data set. In order to help our model learn parameters that generalize better in the future, we add a relative future cost based on the real numbers to the cost presented in Section \ref{sec:error-minim}, so its new formula is introduced in Equation \ref{eq:3}.

\begin{align}
    J(\theta) = \sqrt{\sum_i{(D_{real}(i) - D_{\theta}(i))^2}} + \lambda_{May\, 21} + \lambda_{June\, 3} + \lambda_{June\, 11} \label{eq:3}\\
    \lambda_{date} = 100 \cdot |\frac{R_{date} - P_{date}}{R_{date}}|\\
    R_{date} \equiv Reported\; number\; of\; fatalities\; on\; \textbf{date}\\
    P_{date} \equiv Predicted\; number\; of\; fatalities\; on\; \textbf{date}
\end{align}

We present in Table \ref{optimal-params} the best model parameters found with neural net optimization followed by the refinement step. As stated previously, they minimize the L2 distance between the real and predicted curves of fatalities, from  March 22, 2020, to May 21, 2020, while trying not to overfit the data set by looking ahead until the last available reported date. Several interesting observations are worth making: the estimated basic reproduction number was found to be 2.21, which is very close to the value estimated in the literature. Since it is above 1, it defines an exponential growth in the number of infected people unless reduced by the social distancing measures. The imposed measures of containment indeed reduced the reproduction number by 60\% (from 2.2 to 0.884), so the curve starts decreasing towards zero. The continuous decrease definitely helps our medical staff to manage patients better and safer, which may also explain why the fatality rate found is so small, of only 0.245\%. Note that this value is significantly lower than other numbers reported in the literature so far, which is very good news but it is highly dependent on the number of tested people and on the social dynamics of the observed population.

\begin{table}[H]
  \caption{Best parameters found by our neural net optimization + refinement, in the case of heavy social distancing assumption through the real data.}
  \label{optimal-params}
  \centering
  \begin{tabular}{lll}
    \toprule
    Name & Description & Value \\
    \midrule
    $I_0$ & Initial infectious population & 1725 \\
    $R_0$ & Basic reproduction number & $2.21$ \\
    $T_{inc}$ & Length of incubation period (days) & $2$ \\
    $T_{inf}$ & Length of infectiousness period (days) & $3.47$ \\
    $P_F$ & Case fatality rate & $0.245\%$ \\
    $T_F$ & Time from end of infectiousness to death (days) & $20.3$ \\
    $T_M$ & Recovery time for mild cases (days) & $10$ \\
    $T_V$ & Recovery time for severe cases (days) & $10$ \\
    $P_V$ & Severe symptoms rate & $10\%$ \\
    $P_T$ & Percentage to decrease transmission by after intervention & $60\%$ \\
    $T$ & Intervention time to reduce $R_0$ (days) & $21$ \\
    \bottomrule
  \end{tabular}
\end{table}

Based on this set of parameters, we analyze two cases that influence the daily deaths curve:
\begin{enumerate}
    \item Heavy social distancing, meaning that the enforced norms will not be diminished until the end of the pandemic period;
    \item Moderate social distancing, meaning that on May 15, 2020, the social interaction increased by 10\%.
\end{enumerate}

\subsection{Heavy social distancing}

\begin{figure}[H]
  \centering
  \includegraphics[width=0.8\linewidth]{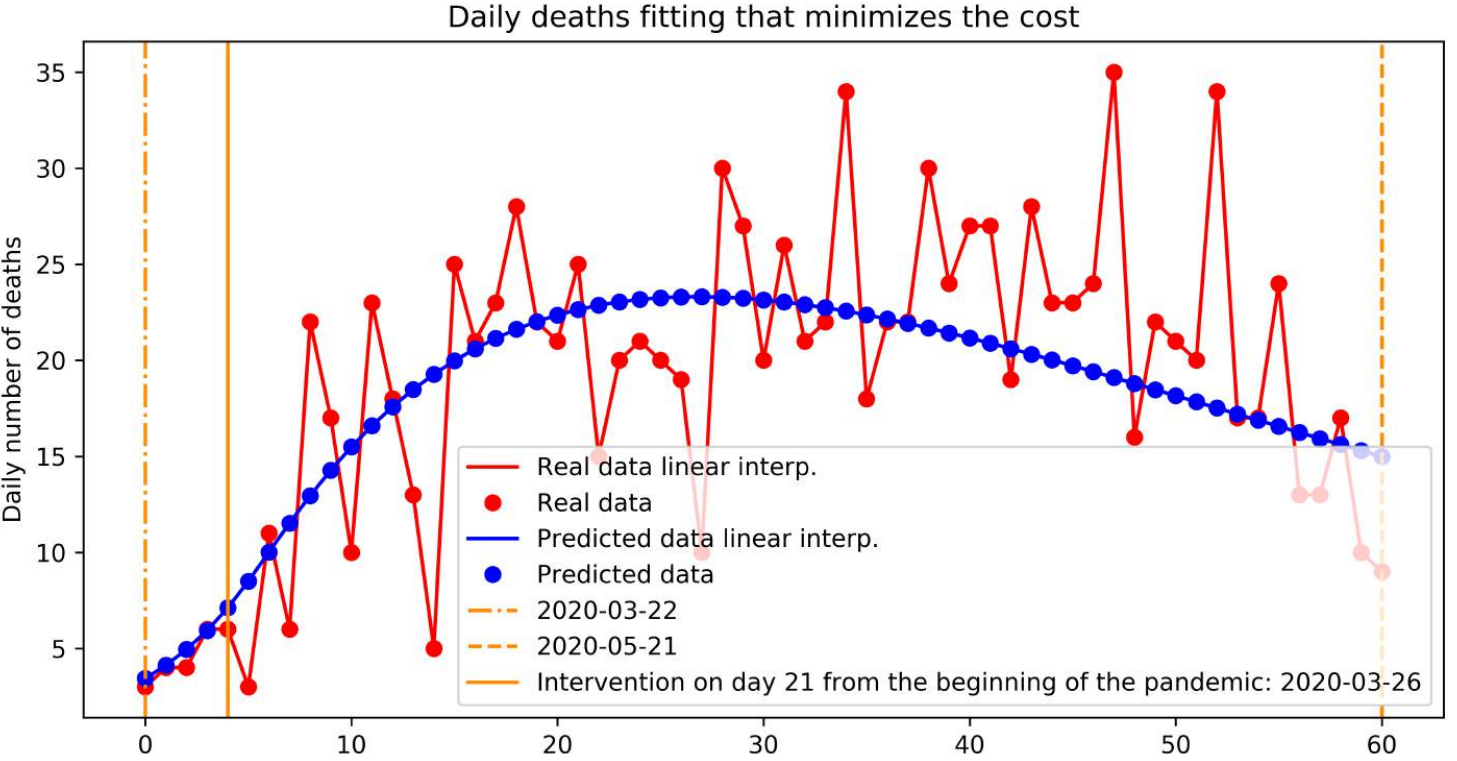}
  \caption{Daily deaths fitted curves. The red dots represent the officially reported daily number of deaths in Romania. The blue line with dots shows our best fit of the Modified-SEIR model through the real data.}
  \label{deaths-fitting}
\end{figure}

Here, we assume that people will adopt a careful behavior and the heavy social distancing rules will apply long after they were proposed ($R_0$ becomes 40\% of its initial value). In Figure \ref{deaths-fitting} we plot our best fitting for the daily deaths approximation using the parameters from Table \ref{optimal-params}. You can notice a change in the convexity of the prediction (blue) curve when the heavy social distancing norms have been adopted (on intervention day).

\begin{figure}[H]
  \centering
  \includegraphics[width=0.8\linewidth]{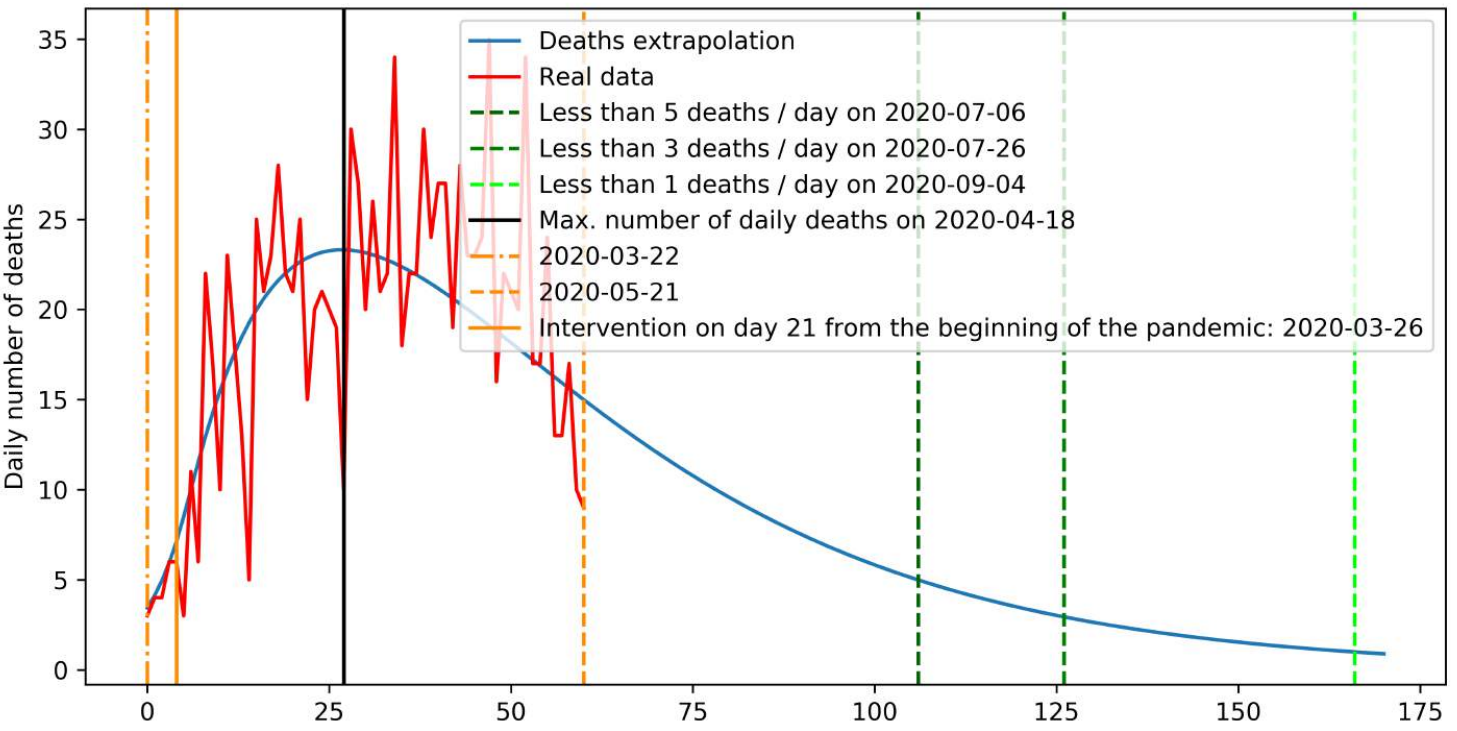}
  \caption{Extrapolation of daily number of fatalities. The red plot represents the officially reported daily number of deaths in Romania. Our extrapolation of daily deaths is the blue line and the green lines represent the days when we estimate less that $5$, $3$ or $1$ daily fatalities.}
  \label{deaths-extrapolation}
\end{figure}

Using the RK4 integrator, as presented in Section \ref{model}, we extrapolate around 200 days from the first day of reported data. In this way, we predict when there are going to be less than $5$, $3$ or $1$ deaths per day, as you can see in Figure \ref{deaths-extrapolation}. A key insight is that the maximum number of daily deaths, the peak of our extrapolation, has already passed on April 18, 2020 (which is right before the Orthodox Easter on April 19-21), meaning that the curve is going to stay under 24 deaths per day.

\begin{figure}[H]
  \centering
  \includegraphics[width=0.8\linewidth]{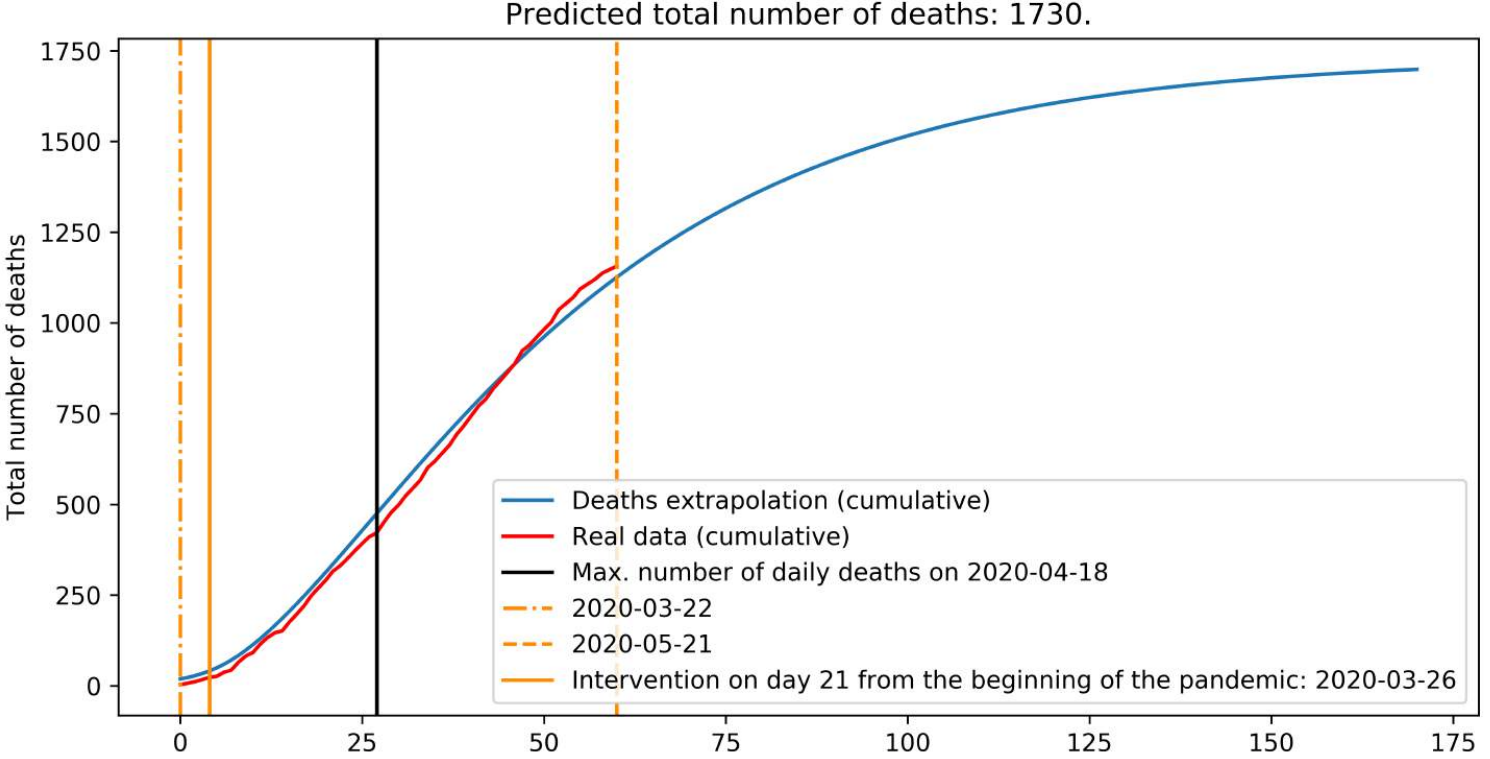}
  \caption{Extrapolation of cumulative fatalities. The red plot represents the officially reported cumulative number of deaths in Romania. Our extrapolation of the cumulative number is the blue line.}
  \label{cum-deaths-extrapolation}
\end{figure}

A subject that is of interest is the predicted total number of fatalities caused by the virus. Using the daily numbers extrapolation, we create a cumulative curve which tells us the total number of fatalities by a certain date. Thus, we predict a total of around 1730 using data from March 22, 2020, to May 21, 2020, as presented in Figure \ref{cum-deaths-extrapolation}.

There is the possibility of optimizing the parameters for fitting the cumulative number of fatalities instead of their daily number. The reason we opted for the latter is three-fold. First, the amount of information that each specific day brings to the cumulative function becomes smaller and smaller each day, converging to 0 at infinity. The cumulative number thus increases with each day and the meaningful information for a given day becomes much smaller than the total cumulated number. It is thus expected that learning could suffer from numerical issues. Second, the number of daily fatalities contains a certain amount of noise that can help us generalize better for our predictions. And third, the curve of daily numbers shows better and clearer when the pandemic peaks and when it is expected to diminish to a non-threatening state.

\begin{figure}[H]
  \centering
  \includegraphics[width=0.8\linewidth]{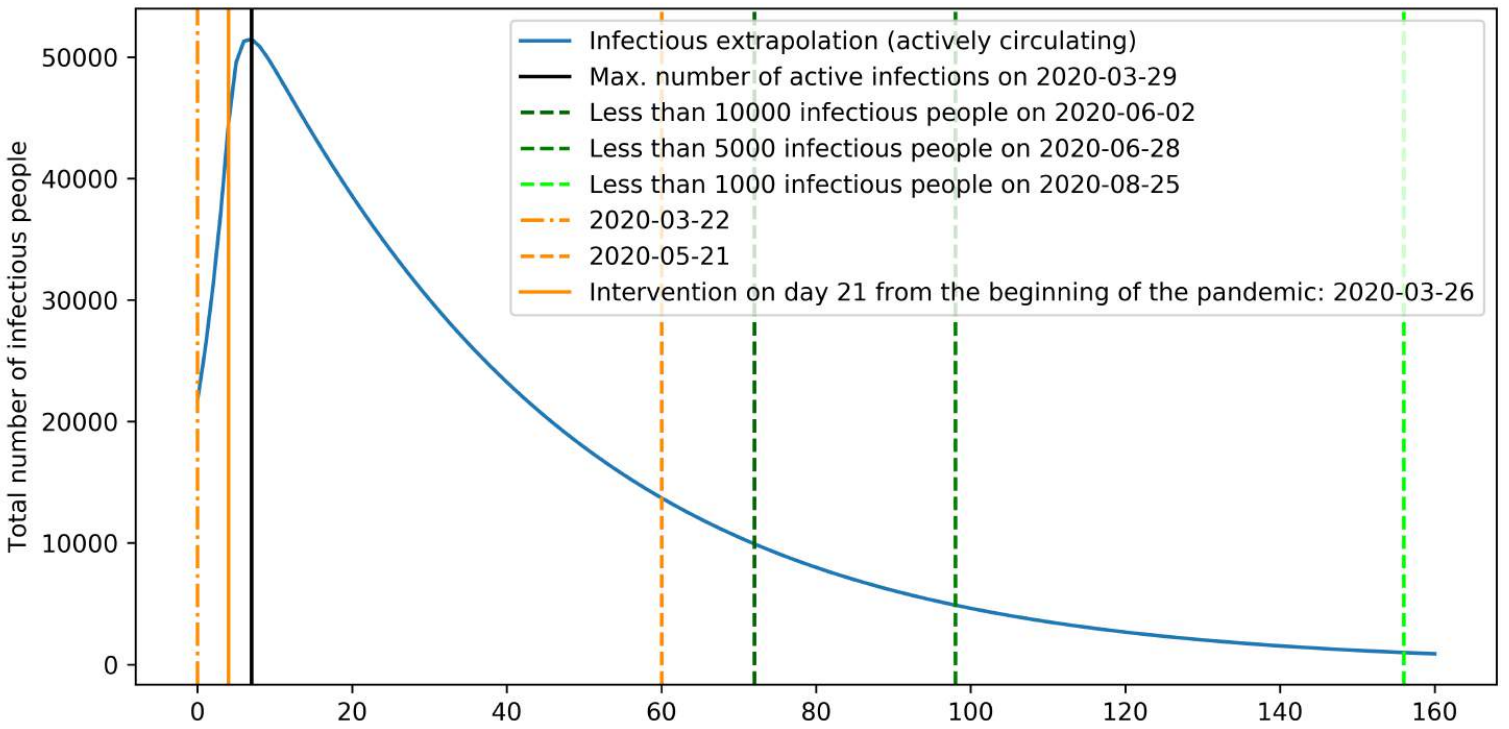}
  \caption{Total number of infectious people extrapolation. We predict less than $10000$ infections on June 2, 2020, less than $5000$ infections on June 8, 2020, and less than $1000$ infections on August 25, 2020.}
  \label{infectious-extrapolation}
\end{figure}

We know that the real number of actively infectious people is hard to obtain, as the number of tested people at a time is just a fraction of the whole population. However, once the right parameters are learned, the model can estimate a total number of infectious people. The results are shown in Figure \ref{infectious-extrapolation}.

\newpage

\subsection{Moderate social distancing}

While we do not know for sure how the social dynamics influence the evolution of the coronavirus, we conduct an experiment assuming that on May 15, 2020, when the social distancing norms became less constraining, the already reduced basic reproduction number (by the heavy social distancing norms) increased by 10\%, while the rest of the parameters remained the same.

\begin{figure}[H]
  \centering
  \includegraphics[width=0.8\linewidth]{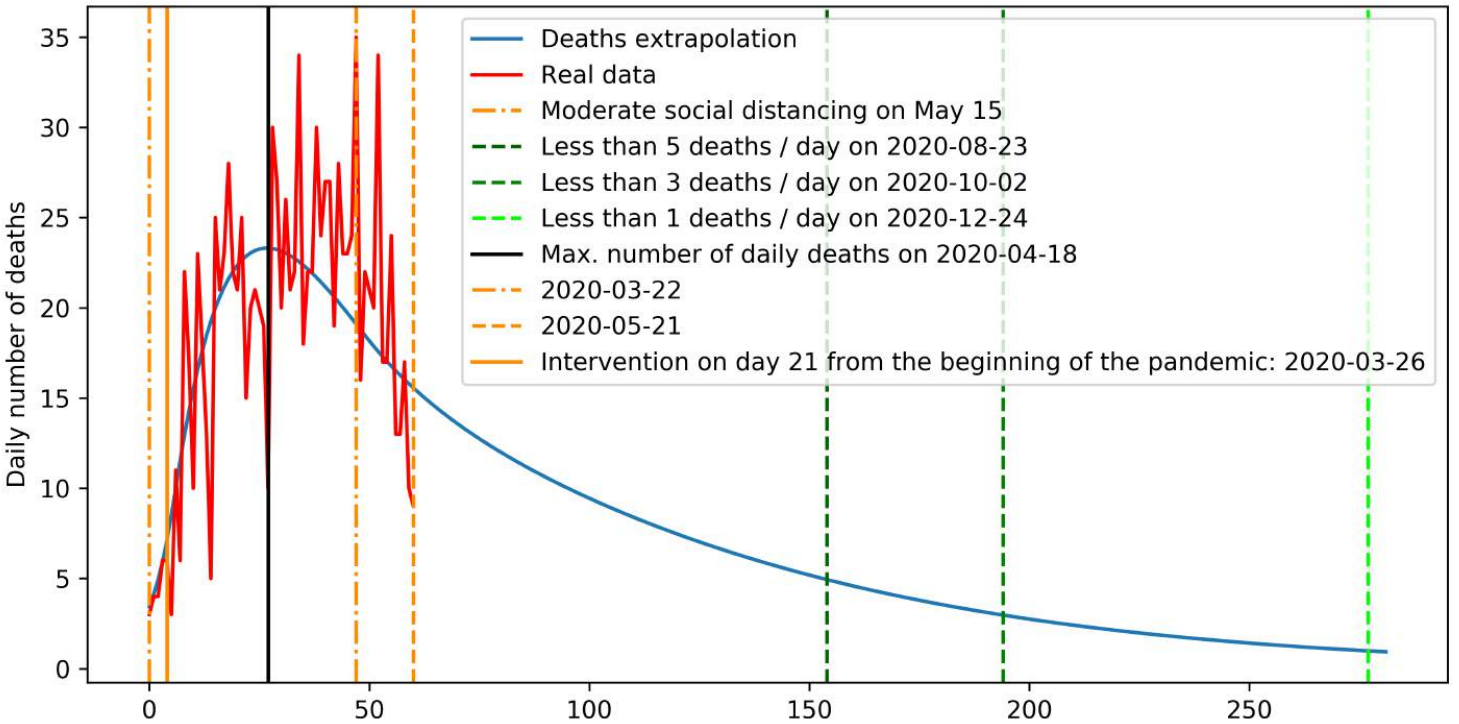}
  \caption{Daily fatalities extrapolation with increased mobility (moderate social distancing) from May 15, 2020. The red plot represents the officially reported daily number of deaths in Romania. Our extrapolation is the blue line and the green lines represent the days when we estimate less that $5$, $3$ or $1$ daily fatalities.}
  \label{deaths-extrapolation-second-wave}
\end{figure}

In Figure \ref{deaths-extrapolation-second-wave} we show in our prediction that the period until there will be less than $5$ daily deaths is prolonged until the end of August, 2020. We do not see a second peak because the basic reproduction number does not get to be over 1 again.

\begin{figure}[H]
  \centering
  \includegraphics[width=0.8\linewidth]{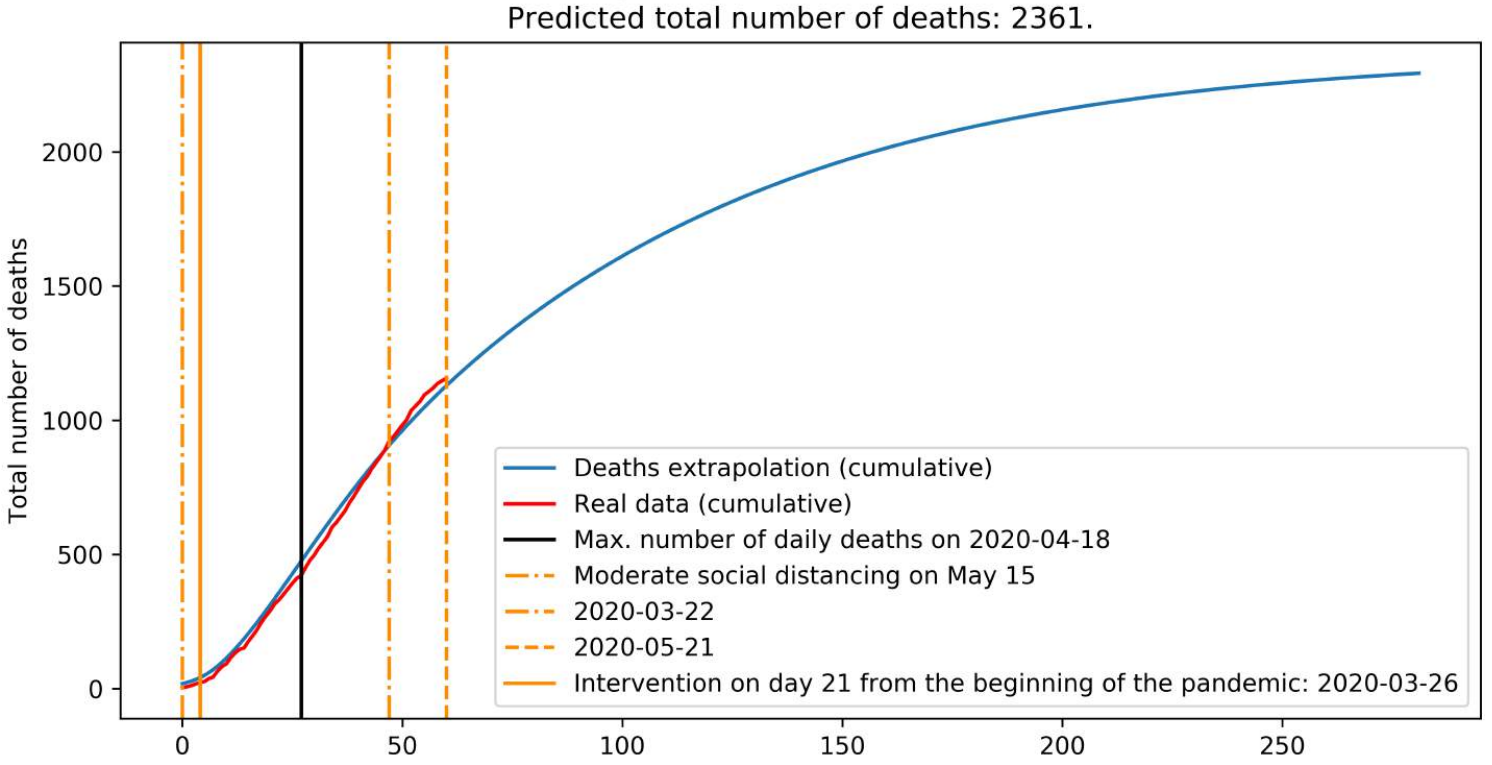}
  \caption{Cumulative deaths extrapolation with increased mobility from May 15, 2020. The red plot represents the officially reported cumulative number of deaths in Romania. Our extrapolation of the cumulative number of deaths is the blue line.}
  \label{cum-deaths-extrapolation-second-wave}
\end{figure}

The total number of deaths jumps to $2361$, shown in Figure \ref{cum-deaths-extrapolation-second-wave}. Compared to \cite{Youyanggu} we seem a bit pessimistic as the number of deaths predicted by us by Aug 1, 2020, is 1861 while their machine learning approach predicts 1776, as of June 12, 2020. Thus, we are still inside their confidence bounds: [1586 - 2187]. Please note that their approach is also data dependent and can offer significantly different predictions by using more data. 

\begin{figure}[H]
  \centering
  \includegraphics[width=0.8\linewidth]{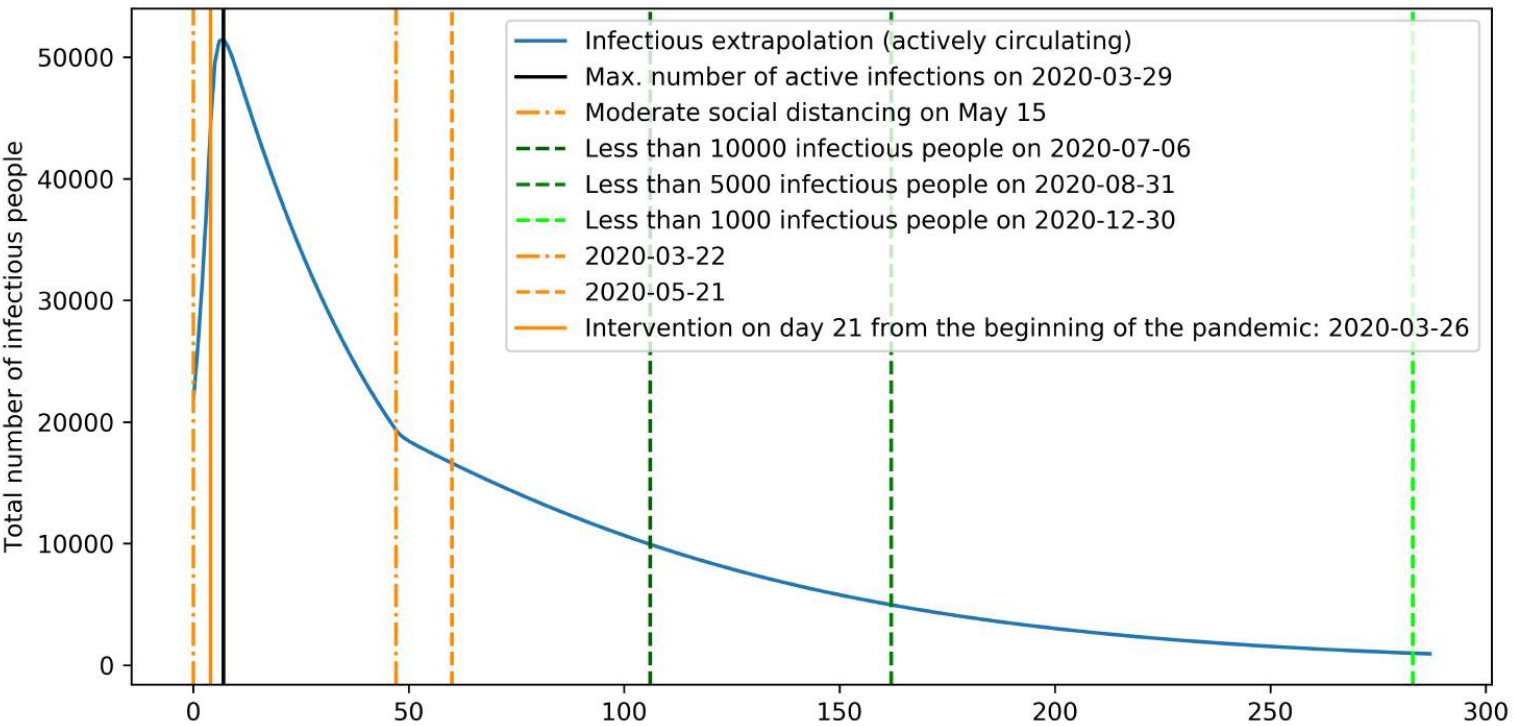}
  \caption{Total number of infectious people extrapolation with less social distancing from May 15, 2020.}
  \label{infectious-extrapolation-second-wave}
\end{figure}

In Figure \ref{infectious-extrapolation-second-wave} one could see the effect of a slight change in the prevention norms, as the total number of actively infected people takes way more time to go down.

\subsection{Validation for our approach}

Ultimately, such modeling is important not only to fit the available observed data and estimate various model parameters, such as the fatality rate, but to predict future events. The ability to predict future outcomes is its real value, in order to prepare the best courses of action in advance. Acting on time, especially in the face of a pandemic, is vital. Thus, in order to test the validity of our model (as it is usually done, in fact, in machine learning) we compare the future predictions made on May 21, 2020 (based on observed data until that date), with the information available until June 11, 2020. Our model is surprisingly accurate with the heavy social distancing assumptions. The predicted total number of fatalities is close to the reported values: 1296 true reported deaths on June 3, 2020, versus 1294 (predicted by the heavy social distancing model) and 1312 (predicted by the moderate social distancing model) and 1369 true reported fatalities on Jun 11, 2020, versus 1375 (predicted by the heavy social distancing model) and 1411 (predicted by the moderate social distancing model).

Regarding the number of infected people, we already know that the number of reported active cases is lower than the real one, so, as mentioned previously, we can use our model to guess the actual number of active infections. Testing our predictions against future data is shown in Table \ref{verify-table}.

\begin{table}[H]
  \caption{Predictions verification. HSD means heavy social distancing and MSD means moderate social distancing.}
  \label{verify-table}
  \centering
  \begin{tabular}{llll}
    \toprule
    Criterion & Reported & HSD prediction & MSD prediction \\
    \midrule
    Total number of deaths on May 3, 2020 & 790 & 807 (2.15\%) & 807 (2.15\%) \\
    Total number of deaths on May 15, 2020 & 1070 & 1031 (3.64\%) & 1031 (-3.64\%) \\
    Total number of deaths on May 21, 2020 & 1156 & 1125 (2.68\%) & 1128 (2.42\%) \\
    Total number of deaths on Jun 3, 2020 & 1296 & 1294 (0.15\%) & 1312 (1.23\%) \\
    Total number of deaths on Jun 11, 2020 & 1369 & 1375 (0.44\%) & 1411 (3.07\%) \\
    Total number of deaths on Jun 19, 2020 & 1484 & 1442 (2.83\%) & 1501 (1.15\%) \\
    
    Active infections on May 3, 2020 & 7504 & 22066 & 22066 \\
    Active infections on May 15, 2020 & 5997 & 16093 & 17673 \\
    Active infections on May 23, 2020 & 5494 & 13000 & 16278 \\
    Active infections on Jun 3, 2020 & 4573 & 9666 & 14478 \\
    Active infections on Jun 11, 2020 & 4530 & 7778 & 13259 \\
    Active infections on Jun 19, 2020 & 5361 & 6253 & 12117 \\
    \bottomrule
  \end{tabular}
\end{table}

\subsection{Predictions for the future}

Now that we have two models, one following heavy social distancing norms and the other following moderate social distancing norms, that have prediction errors < 4\% so far, we can use them as bounds for our future predictions. We summarize our findings in Table \ref{table:predictions}.

\newpage

\begin{table}[H]
  \caption{Predictions. HSD means heavy social distancing and MSD means moderate social distancing.}
  \label{table:predictions}
  \centering
  \begin{tabular}{lll}
    \toprule
    Criterion & HSD prediction & MSD prediction \\
    \midrule
    Total number of deaths on July 1, 2020 & 1521 & 1620\\
    Total number of deaths on August 1, 2020 & 1640 & 1861\\
    Total number of deaths on September 1, 2020 & 1692 & 2027\\
    Total number of deaths on October 1, 2020 & 1713 & 2137\\
    Total number of deaths on November 1, 2020 & 1723 & 2213\\
    Total number of deaths on December 1, 2020 & 1727 & 2262\\
    Total number of deaths & 1730 & 2361 \\
    
    Less than 5 deaths per day & Jul 6, 2020 & Aug 23, 2020 \\
    Less than 3 deaths per day & Jul 26, 2020 & Oct 2, 2020 \\
    Less than 1 deaths per day & Sep 4, 2020 & Dec 24, 2020 \\
    \midrule
    Active infections on July 1, 2020 & 4499 & 10548 \\
    Active infections on August 1, 2020 & 1912 & 7252 \\
    Active infections on September 1, 2020 & 808 & 4900 \\
    Active infections on October 1, 2020 & 351 & 3315 \\
    Active infections on November 1, 2020 & 148 & 2196 \\
    Active infections on December 1, 2020 & 64 & 1467 \\
    
    Less than 10000 active infections & Jun 2, 2020 & Jul 6, 2020\\
    Less than 5000 active infections & Jun 28, 2020 & Aug 31, 2020\\
    Less than 1000 active infections & Aug 25, 2020 & Dec 30, 2020\\
    \bottomrule
  \end{tabular}
\end{table}

\section{Conclusions}

In this paper we propose a computational model to predict the evolution of COVID-19 in Romania and estimate key factors of the pandemics such as the fatality rate, incubation period, infectiousness period and reproduction number, based on the state of the art Modified-SEIR model \cite{chowdhury2020dynamic}. Our technical novelty consists in the way we optimize the parameters of the model, through a self-supervised deep learning approach, in which a convolutional neural network learns from synthetic data, produced by the analytical Modified-SEIR model for random sets of parameters, to predict the correct parameter set - which is known, since it is the one used to generate the synthetic data. Our results show beyond any doubt that our novel self-supervised approach is effective and learning a set of parameters which are not only able to fit the observed data but also to accurately predict in the future, for the three weeks period tested (which is a relatively large period in the case of a rapidly evolving pandemics). 

At the conclusion of our study, we highlight some important findings comprising the total number of fatalities by following the heavy social distancing norms (1730), the total number of deaths following a small decrease in the prevention norms on May 15, 2020, (2361) and the fact that we already passed the peak of the daily number of deaths on April 18, 2020 (one day before the Orthodox Easter). Our predictions are right inside and around the bounds predicted by IHME (1614 deaths by August 1, as of June 12, 2020; \cite{2020.04.21.20074732}) and the ML-based approach presented in Section \ref{related-work} (1776 deaths by August 1, as of June 12, 2020; \cite{Youyanggu}). This and the fact that our set of found parameters are close to the ones presented in the latest literature (for example, an optimal basic reproduction number of 2.21) only empowers the idea that our novel approach can be useful in a fast-paced pandemic, maybe not only for the case of Romania. A notable finding is that the case fatality rate in all the local optima sets seems to be significantly less than 1\%, mostly around 0.245\% and 0.3\% - and this is a optimistic surprise, when compared to the estimates in the literature.

Considering that our model trained on data collected until May 21, 2020, accurately describes the future evolution (future unseen data) of the number of fatalities until June 11, 2020, we conclude that both the model and its inner parameters found provide answers that are very close to the true ones. The results strongly indicate that we should seriously consider data-driven computational approaches, in combination with machine learning, in 
the analysis and decision making process, with respect to fundamental aspects of our lives (such as it is the case of COVID-19 pandemics), for the future and greater good of society.

\section{Declaration}

\subsection{Ethics approval} We did not use any confidential data for the analysis in this paper, and we do not have any ethical issues in this paper.

\subsection{Consent for publication}

We did not use any data that could possibly reveal any personal data of any patient.

\subsection{Availability of data and material} The code used in this paper is available online in a Colab repository: \url{https://colab.research.google.com/drive/1940gRu6cZOhken1ki-PZxeX_VOQTPZ39?usp=sharing}

\subsection{ Competing interests } The authors declare that they have no competing interests.

\subsection{ Funding } There are no funding sources for this paper.

\subsection{Authors' contributions}  All authors contributed equally to this paper.

\bibliographystyle{plainnat}
\bibliography{main.bib}

\begin{thebibliography}{28}
\providecommand{\natexlab}[1]{#1}
\providecommand{\url}[1]{\texttt{#1}}
\expandafter\ifx\csname urlstyle\endcsname\relax
  \providecommand{\doi}[1]{doi: #1}\else
  \providecommand{\doi}{doi: \begingroup \urlstyle{rm}\Url}\fi

\bibitem[Backer et~al.(2020)Backer, Klinkenberg, and Wallinga]{PMID:32046819}
Jantien~A Backer, Don Klinkenberg, and Jacco Wallinga.
\newblock Incubation period of 2019 novel coronavirus (2019-ncov) infections
  among travellers from wuhan, china, 20-28 january 2020.
\newblock \emph{Euro surveillance : bulletin Europeen sur les maladies
  transmissibles = European communicable disease bulletin}, 25\penalty0 (5),
  February 2020.
\newblock ISSN 1025-496X.
\newblock \doi{10.2807/1560-7917.ES.2020.25.5.2000062}.
\newblock URL \url{https://europepmc.org/articles/PMC7014672}.

\bibitem[Chowdhury et~al.(2020)Chowdhury, Heng, Shawon, Goh, Okonofua,
  Ochoa-Rosales, Gonzalez-Jaramillo, Bhuiya, Reidpath, Prathapan,
  et~al.]{chowdhury2020dynamic}
Rajiv Chowdhury, Kevin Heng, Md~Shajedur~Rahman Shawon, Gabriel Goh, Daisy
  Okonofua, Carolina Ochoa-Rosales, Valentina Gonzalez-Jaramillo, Abbas Bhuiya,
  Daniel Reidpath, Shamini Prathapan, et~al.
\newblock Dynamic interventions to control covid-19 pandemic: a multivariate
  prediction modelling study comparing 16 worldwide countries.
\newblock \emph{European journal of epidemiology}, pages 1--11, 2020.

\bibitem[Ciupe and Tuncer(2022)]{Ciupe}
Stanca Ciupe and Necibe Tuncer.
\newblock Identifiability of parameters in mathematical models of sars-cov-2
  infections in humans.
\newblock \emph{Scientific Reports}, 12, 08 2022.
\newblock \doi{10.1038/s41598-022-18683-x}.

\bibitem[Gherghel and Bulai(2020)]{gherghel2020romania}
Iulian Gherghel and Mihai Bulai.
\newblock Is romania ready to face the novel coronavirus (covid-19) outbreak?
  the role of incoming travelers and that of romanian diaspora.
\newblock \emph{Travel Medicine and Infectious Disease}, 2020.

\bibitem[Goh(2020)]{gabgoh}
Gabriel Goh.
\newblock Epidemic calculator, 2020.
\newblock URL \url{https://gabgoh.github.io/COVID/index.html}.

\bibitem[Gu(2020)]{Youyanggu}
Youyang Gu.
\newblock Covid-19 projections using machine learning, 2020.
\newblock URL \url{https://https://covid19-projections.com/}.

\bibitem[Hethcote(2000)]{doi:10.1137/S0036144500371907}
Herbert~W. Hethcote.
\newblock The mathematics of infectious diseases.
\newblock \emph{SIAM Review}, 42\penalty0 (4):\penalty0 599--653, 2000.
\newblock \doi{10.1137/S0036144500371907}.
\newblock URL \url{https://doi.org/10.1137/S0036144500371907}.

\bibitem[IHME COVID-19 health service utilization~forecasting
  team(2020)]{2020.04.21.20074732}
Murray IHME COVID-19 health service utilization~forecasting team,
  Christopher~JL.
\newblock Forecasting the impact of the first wave of the covid-19 pandemic on
  hospital demand and deaths for the usa and european economic area countries.
\newblock \emph{medRxiv}, 2020.
\newblock \doi{10.1101/2020.04.21.20074732}.
\newblock URL
  \url{https://www.medrxiv.org/content/early/2020/04/26/2020.04.21.20074732}.

\bibitem[Kelso et~al.(2009)Kelso, Milne, and Kelly]{socialDistancing}
Joel Kelso, George Milne, and Heath Kelly.
\newblock Simulation suggests that rapid activation of social distancing can
  arrest epidemic development due to a novel strain of influenza.
\newblock \emph{BMC public health}, 9:\penalty0 117, 05 2009.
\newblock \doi{10.1186/1471-2458-9-117}.

\bibitem[Kingma and Ba(2014)]{adam-opti}
Diederik Kingma and Jimmy Ba.
\newblock Adam: A method for stochastic optimization.
\newblock \emph{International Conference on Learning Representations}, 12 2014.

\bibitem[Krivorotko et~al.(2021)Krivorotko, Kabanikhin, Sosnovskaya, and
  Andornaya]{Krivorotko}
Olga Krivorotko, Sergey Kabanikhin, M.~Sosnovskaya, and D.~Andornaya.
\newblock Sensitivity and identifiability analysis of covid-19 pandemic models.
\newblock \emph{Vavilov Journal of Genetics and Breeding}, 25:\penalty0 82--91,
  03 2021.
\newblock \doi{10.18699/VJ21.010}.

\bibitem[Kucharski et~al.(2020)Kucharski, Russell, Diamond, Liu, Edmunds, Funk,
  and Eggo]{Kucharski2020.01.31.20019901}
Adam~J Kucharski, Timothy~W Russell, Charlie Diamond, Yang Liu, John Edmunds,
  Sebastian Funk, and Rosalind~M Eggo.
\newblock Early dynamics of transmission and control of covid-19: a
  mathematical modelling study.
\newblock \emph{medRxiv}, 2020.
\newblock \doi{10.1101/2020.01.31.20019901}.
\newblock URL
  \url{https://www.medrxiv.org/content/early/2020/02/18/2020.01.31.20019901}.

\bibitem[Lau et~al.(2010)Lau, Hsiung, Cowling, Chen, Ho, Tsang, Chang,
  Donnelly, and Leung]{article:lau}
Eric Lau, C~Hsiung, Benjamin Cowling, Chang-Hsun Chen, Lai-Ming Ho, Thomas
  Tsang, Chiu-Wen Chang, Christl Donnelly, and Gabriel Leung.
\newblock A comparative epidemiologic analysis of sars in hong kong, beijing
  and taiwan.
\newblock \emph{BMC infectious diseases}, 10:\penalty0 50, 03 2010.
\newblock \doi{10.1186/1471-2334-10-50}.

\bibitem[Li et~al.(2020)Li, Guan, Wu, Wang, Zhou, Tong, Ren, Leung, Lau, Wong,
  Xing, Xiang, Wu, Li, Chen, Li, Liu, Zhao, Liu, Tu, Chen, Jin, Yang, Wang,
  Zhou, Wang, Liu, Luo, Liu, Shao, Li, Tao, Yang, Deng, Liu, Ma, Zhang, Shi,
  Lam, Wu, Gao, Cowling, Yang, Leung, and Feng]{doi:10.1056/NEJMoa2001316}
Qun Li, Xuhua Guan, Peng Wu, Xiaoye Wang, Lei Zhou, Yeqing Tong, Ruiqi Ren,
  Kathy~S.M. Leung, Eric~H.Y. Lau, Jessica~Y. Wong, Xuesen Xing, Nijuan Xiang,
  Yang Wu, Chao Li, Qi~Chen, Dan Li, Tian Liu, Jing Zhao, Man Liu, Wenxiao Tu,
  Chuding Chen, Lianmei Jin, Rui Yang, Qi~Wang, Suhua Zhou, Rui Wang, Hui Liu,
  Yinbo Luo, Yuan Liu, Ge~Shao, Huan Li, Zhongfa Tao, Yang Yang, Zhiqiang Deng,
  Boxi Liu, Zhitao Ma, Yanping Zhang, Guoqing Shi, Tommy~T.Y. Lam, Joseph~T.
  Wu, George~F. Gao, Benjamin~J. Cowling, Bo~Yang, Gabriel~M. Leung, and Zijian
  Feng.
\newblock Early transmission dynamics in wuhan, china, of novel
  coronavirus–infected pneumonia.
\newblock \emph{New England Journal of Medicine}, 382\penalty0 (13):\penalty0
  1199--1207, 2020.
\newblock \doi{10.1056/NEJMoa2001316}.
\newblock URL \url{https://doi.org/10.1056/NEJMoa2001316}.
\newblock PMID: 31995857.

\bibitem[Long et~al.(2021)Long, Khaliq, and Furati]{Long03082021}
Jie Long, A.~Q.~M. Khaliq, and K.~M. Furati.
\newblock Identification and prediction of time-varying parameters of covid-19
  model: a data-driven deep learning approach.
\newblock \emph{International Journal of Computer Mathematics}, 98\penalty0
  (8):\penalty0 1617--1632, 2021.
\newblock \doi{10.1080/00207160.2021.1929942}.

\bibitem[Marinov and Marinova(2020)]{MARINOV2020100041}
Tchavdar~T. Marinov and Rossitza~S. Marinova.
\newblock Dynamics of covid-19 using inverse problem for coefficient
  identification in sir epidemic models.
\newblock \emph{Chaos, Solitons \& Fractals: X}, 5:\penalty0 100041, 2020.
\newblock ISSN 2590-0544.
\newblock \doi{https://doi.org/10.1016/j.csfx.2020.100041}.
\newblock URL
  \url{https://www.sciencedirect.com/science/article/pii/S2590054420300221}.

\bibitem[Massonis et~al.(2021)Massonis, Banga, and Villaverde]{MASSONIS2021441}
Gemma Massonis, Julio~R. Banga, and Alejandro~F. Villaverde.
\newblock Structural identifiability and observability of compartmental models
  of the covid-19 pandemic.
\newblock \emph{Annual Reviews in Control}, 51:\penalty0 441--459, 2021.
\newblock ISSN 1367-5788.
\newblock \doi{https://doi.org/10.1016/j.arcontrol.2020.12.001}.
\newblock URL
  \url{https://www.sciencedirect.com/science/article/pii/S1367578820300778}.

\bibitem[Mihaela van~der Schaar(2020)]{how-artificial}
Ahmed~Alaa Mihaela van~der Schaar.
\newblock How artificial intelligence and machine learning can help healthcare
  systems respond to covid-19.
\newblock 3 2020.

\bibitem[Petrica et~al.(2022)Petrica, Stochitoiu, Leordeanu, and
  Popescu]{petrica2022regime}
Marian Petrica, Radu~D Stochitoiu, Marius Leordeanu, and Ionel Popescu.
\newblock A regime switch analysis on covid-19 in romania.
\newblock \emph{Scientific Reports}, 12\penalty0 (1):\penalty0 15378, 2022.

\bibitem[Popescu et~al.(2020)Popescu, Marin, Melinte, Gherlan, Banicioiu,
  Dogaru, Smadu, Veja, Nedu, Stanciu, et~al.]{popescu2020covid}
Corneliu~Petru Popescu, Alexandru Marin, Violeta Melinte, George~Sebastian
  Gherlan, Filofteia~Cojanu Banicioiu, Adelina Dogaru, Sebastian Smadu,
  Ana~Maria Veja, Elena Nedu, Delia Stanciu, et~al.
\newblock Covid-19 in a tertiary hospital from romania: Epidemiology,
  preparedness and clinical challenges.
\newblock \emph{Travel Medicine and Infectious Disease}, 2020.

\bibitem[Qian et~al.(2020)Qian, Alaa, and Schaar]{when-to-lift}
Zhaozhi Qian, Ahmed Alaa, and Mihaela Schaar.
\newblock When to lift the lockdown? global covid-19 scenario planning and
  policy effects using compartmental gaussian processes.
\newblock 05 2020.

\bibitem[Read et~al.(2020)Read, Bridgen, Cummings, Ho, and
  Jewell]{Read2020.01.23.20018549}
Jonathan~M Read, Jessica~RE Bridgen, Derek~AT Cummings, Antonia Ho, and Chris~P
  Jewell.
\newblock Novel coronavirus 2019-ncov: early estimation of epidemiological
  parameters and epidemic predictions.
\newblock \emph{medRxiv}, 2020.
\newblock \doi{10.1101/2020.01.23.20018549}.
\newblock URL
  \url{https://www.medrxiv.org/content/early/2020/01/28/2020.01.23.20018549}.

\bibitem[Sikder et~al.(2023)Sikder, Hossain, and Islam]{Sikder}
Arun Sikder, Md.~Biplob Hossain, and Md~Hamidul Islam.
\newblock Compartmental modelling in epidemic diseases: a comparison between
  sir model with constant and time-dependent parameters.
\newblock \emph{Inverse Problems}, 39, 02 2023.
\newblock \doi{10.1088/1361-6420/acb4e7}.

\bibitem[Tan~Delin(2012)]{Tan2012}
Chen~Zheng Tan~Delin.
\newblock On a general formula of fourth order runge-kutta.
\newblock \emph{Journal of Mathematical Science \& Mathematics Education},
  2012.
\newblock URL \url{http://w.msme.us/2012-2-1.pdf}.

\bibitem[Virlogeux et~al.(2016)Virlogeux, Fang, Park, Wu, and
  Cowling]{article:virlogeux}
Victor Virlogeux, Vicky Fang, Minah Park, Jianhong Wu, and Benjamin Cowling.
\newblock Comparison of incubation period distribution of human infections with
  mers-cov in south korea and saudi arabia.
\newblock \emph{Scientific Reports}, 6, 10 2016.
\newblock \doi{10.1038/srep35839}.

\bibitem[WHO(2020)]{who-report}
World Health~Organization WHO.
\newblock Report of the who-china joint mission on coronavirus disease 2019
  (covid-19), 2020.
\newblock URL
  \url{https://www.who.int/publications-detail/report-of-the-who-china-joint-mission-on-coronavirus-disease-2019-(covid-19)}.

\bibitem[Wright(2015)]{doi:10.1007/s10107-015-0892-3}
Stephen~J. Wright.
\newblock Coordinate descent algorithms.
\newblock \emph{Mathematical Programming}, 2015.
\newblock \doi{10.1007/s10107-015-0892-3}.
\newblock URL \url{https://arxiv.org/abs/1502.04759}.

\bibitem[Wu et~al.(2020)Wu, Leung, and Leung]{WU2020689}
Joseph~T Wu, Kathy Leung, and Gabriel~M Leung.
\newblock Nowcasting and forecasting the potential domestic and international
  spread of the 2019-ncov outbreak originating in wuhan, china: a modelling
  study.
\newblock \emph{The Lancet}, 395\penalty0 (10225):\penalty0 689 -- 697, 2020.
\newblock ISSN 0140-6736.
\newblock \doi{https://doi.org/10.1016/S0140-6736(20)30260-9}.
\newblock URL
  \url{http://www.sciencedirect.com/science/article/pii/S0140673620302609}.

\end{thebibliography}

\section{Appendix}
\setcounter{equation}{0}
The equations of the proposed model are: 

\begin{table}[H]
  \centering
  \renewcommand{\arraystretch}{2}
  \begin{tabular}{ll}
    $\dv{S}{t} = - \beta I S$ & $\dv{E}{t} = \beta I S - \sigma E$ \\[10pt]
    
    $\dv{I}{t} = \sigma E - \gamma I$ & $\dv{M}{t} = P_M \gamma I - \frac{1}{T_M} M$ \\[10pt]

    $\dv{V}{t} = P_V \gamma I - \frac{1}{T_H} V$ & $\dv{H}{t} = \frac{1}{T_H} V - \frac{1}{T_V} H$ \\[10pt]

    $\dv{F}{t} = P_F \gamma I - \frac{1}{T_F} F$ & $\dv{R_M}{t} = \frac{1}{T_M} M$ \\[10pt]

    $\dv{R_V}{t} = \frac{1}{T_V} H$ & $\dv{R_F}{t} = \frac{1}{T_F} F$ \\[10pt]
    
    $\beta = 
    \begin{cases}
        \frac{R_0}{T_{inf}}, & \text{before } T\\
        \frac{(1-P_T) R_0}{T_{inf}}, & \text{otherwise}
    \end{cases}$ & $\sigma = \frac{1}{T_{inc}}$ \\
    $\gamma = \frac{1}{T_{inf}}$ & $P_M = 1 - P_V - P_F$ \\
  \end{tabular}
\end{table}

We point out that the parameters $(\beta, \sigma, \gamma, P_M, P_V, P_F, T_M, T_H, T_V, T_F)$ are positive constants.

\subsection{Identification of the steady states}

A steady state occurs when all time derivatives are zero.

\begin{itemize}
    \item From \(\frac{dR_M}{dt} = \frac{1}{T_M} M = 0\), since \(T_M > 0\), we have \(M = 0\).
    \item From \(\frac{dR_V}{dt} = \frac{1}{T_V} H = 0\), since \(T_V > 0\), we have \(H = 0\).
    \item From \(\frac{dR_F}{dt} = \frac{1}{T_F} F = 0\), since \(T_F > 0\), we have \(F = 0\).
    \item From \(\frac{dM}{dt} = P_M \gamma I - \frac{1}{T_M} M = 0\), with \(M = 0\), we get \(P_M \gamma I = 0\). Since \(P_M, \gamma > 0\), we have \(I = 0\).
    \item From \(\frac{dI}{dt} = \sigma E - \gamma I = 0\), with \(I = 0\), \(\sigma E = 0\). Since \(\sigma > 0\), we have \(E = 0\).
    \item From \(\frac{dS}{dt} = -\beta I S = 0\), with \(I = 0\), \(S\) is unconstrained.
    \item From \(\frac{dE}{dt} = \beta I S - \sigma E = 0\), with \(I = 0\), \(E = 0\), this is satisfied.
    \item From \(\frac{dV}{dt} = P_V \gamma I - \frac{1}{T_H} V = 0\), with \(I = 0\), \(\frac{1}{T_H} V = 0\). Since \(T_H > 0\), we have \(V = 0\).
    \item From \(\frac{dH}{dt} = \frac{1}{T_H} V - \frac{1}{T_V} H = 0\), with \(V = 0\), \(H = 0\), this is satisfied.
    \item From \(\frac{dF}{dt} = P_F \gamma I - \frac{1}{T_F} F = 0\), with \(I = 0\), \(F = 0\), this is satisfied.
\end{itemize}

Now, consider the equations for $R_M$, $R_V$, and $R_F$:

\begin{itemize}
\item $\frac{dR_M}{dt} = \frac{1}{T_M} M = 0$ since $M = 0$, so $R_M$ is constant (denote it by $R_M^*$).
\item $\frac{dR_V}{dt} = \frac{1}{T_V} H = 0$ since $H = 0$, so $R_V$ is constant (denote it by $R_V^*$).
\item $\frac{dR_F}{dt} = \frac{1}{T_F} F = 0$ since $F = 0$, so $R_F$ is constant (denote it by $R_F^*$).
\end{itemize}

Finally, equation \ref{eq:1} ($-\beta I S = 0$) with $I = 0$ imposes no constraint on $S$, suggesting $S$ can be any non-negative value (since $S$ likely represents a population, $S \geq 0$), say $S^*$.

Thus, the steady states are:
$$(S, E, I, M, V, H, F, R_M, R_V, R_F) = (S^*, 0, 0, 0, 0, 0, 0, R_M^*, R_V^*, R_F^*)$$
where $S^* \geq 0$ and $R_M^*$, $R_V^*$, $R_F^*$ are arbitrary constants. 

We should also notice that the sum of all variables
\begin{equation}\label{e:param1}
S+E+I+M+V+H+F+R_M+R_V+R_F=N
\end{equation}
remains constant in time and equals the total population at time $0$.  Thus we also obtain that 
\[
S^*+R_M^*+R_V^*+R_F^*=N.
\]

The three recovered compartments incorporate the mass that has accumulated up to the moment the infection dies out.

\subsection{Jacobian at a disease-free equilibrium}

To find the eigenvalues, we compute the Jacobian matrix of the system at the steady state. Define the state vector as $(S, E, I, M, V, H, F, R_M, R_V, R_F)$, and the system as $\frac{d\mathbf{x}}{dt} = \mathbf{f}(\mathbf{x})$, where:

$$\mathbf{f} = \left( -\beta I S, \beta I S - \sigma E, \sigma E - \gamma I, P_M \gamma I - \frac{1}{T_M} M, P_V \gamma I - \frac{1}{T_H} V, \frac{1}{T_H} V - \frac{1}{T_V} H, P_F \gamma I - \frac{1}{T_F} F, \frac{1}{T_M} M, \frac{1}{T_V} H, \frac{1}{T_F} F \right)$$
The Jacobian $J = \frac{\partial \mathbf{f}}{\partial \mathbf{x}}$ is a 10×10 matrix. We evaluate it at the steady state $(S^*, 0, 0, 0, 0, 0, 0, R_M^*, R_V^*, R_F^*)$.

With the above ordering of variables, the Jacobian \(J\bigl(S_0\bigr)=\partial f/\partial x\) evaluated at any disease-free equilibrium is block--lower-triangular:

\[
J = \begin{pmatrix}
0 & 0 & -\beta S^* & 0 & 0 & 0 & 0 & 0 & 0 & 0 \\
0 & -\sigma & \beta S^* & 0 & 0 & 0 & 0 & 0 & 0 & 0 \\
0 & \sigma & -\gamma & 0 & 0 & 0 & 0 & 0 & 0 & 0 \\
0 & 0 & P_M \gamma & -\frac{1}{T_M} & 0 & 0 & 0 & 0 & 0 & 0 \\
0 & 0 & P_V \gamma & 0 & -\frac{1}{T_H} & 0 & 0 & 0 & 0 & 0 \\
0 & 0 & 0 & 0 & \frac{1}{T_H} & -\frac{1}{T_V} & 0 & 0 & 0 & 0 \\
0 & 0 & P_F \gamma & 0 & 0 & 0 & -\frac{1}{T_F} & 0 & 0 & 0 \\
0 & 0 & 0 & \frac{1}{T_M} & 0 & 0 & 0 & 0 & 0 & 0 \\
0 & 0 & 0 & 0 & 0 & \frac{1}{T_V} & 0 & 0 & 0 & 0 \\
0 & 0 & 0 & 0 & 0 & 0 & \frac{1}{T_F} & 0 & 0 & 0
\end{pmatrix}.
\]
Since $R_M$, $R_V$, and $R_F$ do not appear in the equations for other variables, the Jacobian is block triangular:
$$J = \begin{pmatrix}
A & 0 \\
B & D
\end{pmatrix}$$
\begin{itemize}
\item $A$: 7×7 submatrix for $(S, E, I, M, V, H, F)$,
\item $D$: 3×3 diagonal submatrix for $(R_M, R_V, R_F)$ with entries $0, 0, 0$,
\item $B$: coupling terms,
\item $0$: zero matrix.
\end{itemize}

The eigenvalues of $J$ are the eigenvalues of $A$ union those of $D$. Since $D = \text{diag}(0, 0, 0)$, it has three zero eigenvalues.
For $A$:
$$A = \begin{pmatrix}
0 & 0 & -\beta S^* & 0 & 0 & 0 & 0 \\
0 & -\sigma & \beta S^* & 0 & 0 & 0 & 0 \\
0 & \sigma & -\gamma & 0 & 0 & 0 & 0 \\
0 & 0 & P_M \gamma & -\frac{1}{T_M} & 0 & 0 & 0 \\
0 & 0 & P_V \gamma & 0 & -\frac{1}{T_H} & 0 & 0 \\
0 & 0 & 0 & 0 & \frac{1}{T_H} & -\frac{1}{_clsT_V} & 0 \\
0 & 0 & P_F \gamma & 0 & 0 & 0 & -\frac{1}{T_F}
\end{pmatrix}$$

$A$ is block triangular:

\begin{itemize}

\item First block: $A_1 = \begin{pmatrix} 0 & 0 & -\beta S^* \\ 0 & -\sigma & \beta S^* \\ 0 & \sigma & -\gamma \end{pmatrix}$,
\item Second block: $-\frac{1}{T_M}$,
\item Third block: $\begin{pmatrix} -\frac{1}{T_H} & 0 \\ \frac{1}{T_H} & -\frac{1}{T_V} \end{pmatrix}$,
\item Fourth block: $-\frac{1}{T_F}$.
\end{itemize}

The eigenvalues of $A$ are the union of the eigenvalues of these blocks.
\begin{itemize}
    \item The first block, $A_1$:
$$A_1 - \lambda I = \begin{pmatrix}
-\lambda & 0 & -\beta S^* \\
0 & -\sigma - \lambda & \beta S^* \\
0 & \sigma & -\gamma - \lambda
\end{pmatrix}$$

The characteristic polynomial is:  
$$(-\lambda) [ (-\sigma - \lambda)(-\gamma - \lambda) - (\beta S^*)(\sigma) ] = (-\lambda) [ \lambda^2 + (\sigma + \gamma) \lambda + \sigma (\gamma - \beta S^*) ] = 0$$

The roots of the characteristic polynomial are
$$\lambda_1 = 0$$
$$\lambda_{2,3} = \frac{-(\sigma + \gamma) \pm \sqrt{(\sigma + \gamma)^2 - 4 \sigma (\gamma - \beta S^*)}}{2}$$

\item The second block, for $-\frac{1}{T_M}$, gives:
$$\lambda = -\frac{1}{T_M}$$

\item The third block, $\begin{pmatrix} -\frac{1}{T_H} & 0 \\ \frac{1}{T_H} & -\frac{1}{T_V} \end{pmatrix}$, has the characteristic polynomial:
$$\det \begin{pmatrix} -\frac{1}{T_H} - \lambda & 0 \\ \frac{1}{T_H} & -\frac{1}{T_V} - \lambda \end{pmatrix} = (-\frac{1}{T_H} - \lambda)(-\frac{1}{T_V} - \lambda) = 0$$
which gives the solutions
$$\lambda = -\frac{1}{T_H}$$
$$\lambda = -\frac{1}{T_V}$$

\item The fourth block, For $-\frac{1}{T_F}$, gives 
$$\lambda = -\frac{1}{T_F}$$
\end{itemize}

Hence, the eigenvalues of $J$ are:

\[
\boxed{
\begin{aligned}
&\lambda = 0 \quad \text{(multiplicity 4: } S,\, R_M,\, R_V,\, R_F),\\
&\lambda_{2,3} = -\frac{\gamma + \sigma}{2} \pm \frac{1}{2} \sqrt{(\gamma - \sigma)^2 + 4 \beta \sigma S^*},\\
&\lambda = -\frac{1}{T_M},\; -\frac{1}{T_F},\; -\frac{1}{T_H},\; -\frac{1}{T_V}.
\end{aligned}
}
\]

\subsection{Interpretation}

The system admits many equilibrium points; in fact, the set of critical points forms a 3-dimensional manifold, parameterized by the four “susceptible/recovered” variables.

$$(S^*,\,R_M^*,\,R_V^*,\,R_F^*)$$

subject to the overall population constraint
$$\;S^* + R_M^* + R_V^* + R_F^* + E^* + I^* + H^* = N.$$

Linearizing around any such point, the Jacobian has:

\begin{enumerate}
    \item \textbf{Four zero eigenvalues} corresponding to $(S,R_M,R_V,R_F)$. These modes are neutral (rate 0), reflecting the fact that once on the manifold of \emph{susceptible} and \emph{recovered}, there is no restoring force to push it away from it.
    \item \textbf{Four strictly negative eigenvalues} for the fast compartments $(M,V,H,F)$, implying these populations relax exponentially quickly onto the slow manifold.
    \item \textbf{Two eigenvalues} $\lambda_{2,3}$ governing $(E,I)$, with

 $$\lambda_2 + \lambda_3 = -(\sigma + \gamma)\quad\text{and}\quad\lambda_2\,\lambda_3 = \sigma\,(\gamma - \beta\,S^*). $$
\end{enumerate}

\subsection{Stability of the model}

This section is dedicated to proving that the main system of the paper is indeed globally stable.  Here is the formal result.  

\begin{theorem}
The system \ref{seir-eqs-table} is globally stable for the parameters 
\[
\beta, \sigma, \gamma, P_M, P_V, P_F, T_M, T_H, T_V, T_F >0
\]
and the initial conditions 
\[
S(0)>0, I(0), E(0), M(0), V(0), H(0),  F(0), R_M(0), R_V(0), R_F(0) \ge0,
\] with at least $I(0)>0$ or $E(0)>0$.  
\end{theorem}

\begin{proof}

Let us have a closer look at the subsystem corresponding to $(S,E,I)$, being the system responsible for the stability of the model:

\begin{equation} \label{sei}
  \left\{
    \begin{aligned}
      & \dv{S}{t} = - \beta I S \\
      & \dv{E}{t} = \beta I S - \sigma E\\
      & \dv{I}{t} = \sigma E - \gamma I\\
      &  \text{with }  E(0)\geq 0, I(0) \geq 0.
    \end{aligned}
  \right.
\end{equation}

The subsystem is locally Lipschitz on the positive region 
$\{(S,E,I)\in\mathbb R^3 : S\ge0,\;E\ge0,\;I\ge0\}
$, so solutions can be continued as long as they stay nonnegative and do not explode. 

Knowing that $E\geq 0$, $I\geq 0$, $S\geq 0$ and $S(t)=S_0 e^{-\beta \int_{0}^tI(s)ds}$, we obtain $S(t)>0$ for $I(s)$ well defined. 
From the initial conditions, we have $E(0)\geq 0$ and $I(0) \geq 0$ with $E(0)> 0$ or $I(0) > 0$.

Now, take a time $t_0$ such that  $E(t_0)\geq 0$ or $I(t_0)\ge 0$. Then
\[
e^{\sigma t}E(t)=e^{\sigma t_0}E(t_0)+\int_{t_0}^t\beta S(u) I(u)du
\]
which implies that for $\epsilon>0$ and $t\leq t_0+ \epsilon$ the inequality $E(t)>0$ holds. Furthermore. 
\[
e^{\gamma t}I(t)=e^{\gamma t_0}I(t_0)+\sigma \int_{t_0}^t e^ {\gamma u}E(u)du >0,\text{ for }t\in[t_0,t_0+\epsilon]. 
\]
As a remark, as long as the solution exists and $I(0)\ge0$ and $E(0)\ge0$ with at least one of them being positive, for the whole time $t\ge0$ we will have $E(t)>0$ and $I(t)>0$.\\  

Using the summation of the equations of the subsystem \ref{sei}, we obtain:
\begin{equation*}
    \dv{}{t} {(S(t)+E(t)+I(t))=-\gamma I(t)  \leq 0}
\end{equation*}

So, $S(t)+E(t)+I(t)\leq S(0)+E(0)+I(0)$. In particular, this also implies that all quantities $S(t), E(t), I(t)$ are bounded for all $t\ge0$, and this implies that the solution is defined for all $t\ge0$.  

Next, $S(t)=S(0)e^{-\beta\int_{0}^t I(u)du}$ and we can notice that $S(t)$ is a decreasing function; hence, \[
S(t)\xrightarrow{t\rightarrow\infty} S^*.
\]

We know that $S(t)+E(t)+I(t)$ is also decreasing, thus $S(t)+E(t)+I(t)$ is convergent when $t\rightarrow\infty$ and automatically we can deduce that $E(t)+I(t)$ is convergent when $t\rightarrow\infty$.\\

Let us observe now that:
$$e^{\gamma t}I(t)=e^{\gamma t_0}I(t_0)+\int_{t_0}^{t} \sigma e^{\gamma u}E(u)du\geq e^{\gamma t_0}I(t_0)$$

Thus $e^{\gamma t_0}I(t_0)\geq \epsilon$ and for $t\leq t_0+\frac{\ln(2)}{\gamma}$ we have $I(t)\geq \frac{\epsilon}{2}$. So, there exists $\epsilon>0 $ and a sequence $t_n\xrightarrow[n\rightarrow\infty]{} \infty$ with $t_{n+1}-t_n \rightarrow\infty$, such as $I(t_n)\geq \epsilon$. By taking $t-t_n\leq \frac{\ln(2)}{\gamma}$ we have that $I(t)\geq \frac{\epsilon}{2}$.\\

Therefore, 

\begin{align*}
    S(t)+E(t)+I(t) &= S(t_n)+E(t_n)+I(t_n)-\gamma \int_{t_n}^t I(u)du\\
    &\leq S(t_n)+E(t_n)+I(t_n)-\gamma (t-t_n)\frac{\epsilon}{2}
\end{align*}

For $t=\bar{t}_n=t_n+\frac{\ln(2)}{\gamma}$, we have:
$$S(\bar{t}_n)+E(\bar{t}_n)+I(\bar{t}_n)\leq S(t_n)+E(t_n)+I(t_n)-\frac{\epsilon \cdot \ln(2)}{2}.$$

So, because $t_{n+1}-t_n \rightarrow \infty$ we have
$$0\leq S(t_{n+1})+E(t_{n+1})+I(t_{n+1})\leq S(t_n)+E(t_n)+I(t_n)-\frac{\epsilon \cdot \ln(2)}{2}$$ that results in $S(t_n)+E(t_n)+I(t_n) \rightarrow -\infty$, which is a contradiction. Hence, $I(t) \xrightarrow{t \rightarrow\infty} 0$ \\

Using the second equation of the system \ref{sei}, we have
$$e^{\sigma t}E(t) = e^{\sigma t_0} E(t_o)+\beta \int_{t_0}^{t} e^{\sigma u} I(u)S(u) du.$$ 
For $I(t) \xrightarrow{t \rightarrow\infty} 0$ and knowing $0\leq S(u)\leq S(0)+E(0)+I(0)$ we obtain 
\[
0\leq e^{\sigma t} E(t)\leq e^{\sigma t_{\epsilon}}E(t_{\epsilon})+\beta \epsilon(S(0)+E(0)+I(0))(e^{\sigma t}-e^{\sigma t_{\epsilon}}), \text{ for } t\geq t_{\epsilon} \text{ with } |I(t)|\leq \epsilon.  
\]
This concludes that $E(t)\rightarrow 0$.\\

Next, from the first equation of the system \ref{sei}, we obtain
\[
S(t)=S_0 e^{-\beta \int_0^t I(u) du}.
\]

At the same time $S(t)+E(t)+I(t)=S(0)+E(0)+I(0)-\gamma \int_o^tI(u)du$, so $$S(t)=S(0)e^{\frac{\beta}{\gamma}(S(t)+E(t)+I(t)-S(0)-E(0)-I(0))}$$
from which we deduce that 
$$\frac{\beta}{\gamma}S(t)e^{-\frac{\beta}{\gamma}S(t)}=\frac{\beta}{\gamma}S(0)e^{-\frac{\beta}{\gamma}(E(t)+I(t)-S(0)-E(0)-I(0))}$$
obtaining, in the limit, 
$$
S^*=S_0e^{\frac{\beta}{\gamma}(S^*-S(0)-E(0)-I(0))}.
$$

Combining the last two equations of the system \ref{sei} gives:

\begin{equation*}
    \dv{}{t} {(E(t)+I(t))=(\beta S(t)-\gamma)I}
\end{equation*}

We know that $\beta S(t)-\gamma \rightarrow \beta S^*-\gamma $. If $\beta S^*-\gamma \geq 0$ then $\beta S(t)-\gamma\geq 0$ because $S(t)\searrow S^*$, when $t \rightarrow \infty$. 
Hence, $\dv{}{t} {(E(t)+I(t))\geq 0}$ and automatically $E(t)+I(t)\geq E(0)+I(0)>0$ if $E(0)+I(0)>0$, from which we conclude that $E(t)+I(t)$ does not converge to $0$. The conclusion of this is that 
\[
\beta S^*<\gamma
\]
and that, in regards to the stability of the main system \ref{seir-eqs-table} implies that the system is stable. \\

Thus, so far we proved that $S(t)$ converges, $E(t)$ and $I(t)$ converge to $0$. Next, we will prove that the other quantities of the initial system also converge.

Let us consider a general equation:

\begin{equation*}
    \dv{W}{t} = aI-bW
\end{equation*}
with $a>0, b>0$ and $W(0)\ge0$, then $W(t)$ converges to $0$.  Indeed, notice that  
\[
e^{bt}W(t)-a\int_0^t e^{bu}I(u)du=W(0), \text{ for all }t>0.  
\]
Then, in the first place $W(t)\ge0$ and moreover, 
\begin{equation}\label{eqm}
    W(t)=e^{-bt}W(0)+ae^{-bt}\int_0^t e^{bu}I(u)du.
\end{equation}

If $I(t) \xrightarrow{t \rightarrow\infty} 0$ we have:
\begin{align}
\left|e^{-bt}\int_0^t e^{bu}I(u)du \right| 
&\leq \left| e^{-bt}\int_0^{t_{\epsilon}} e^{bu}I(u)du\right| + e^{-bt}\int_0^t \epsilon \cdot e^{bu}du\nonumber\\
&\leq e^{-bt}\int_0^{t_{\epsilon}} e^{bu}|I(u)|du+\frac{\epsilon(1-e^{-bt})}{b}\nonumber
\end{align}
for $|I(t)|\leq \epsilon, \forall t\geq t_{\epsilon}$. Therefore, we can conclude that  $W(t) \xrightarrow{t \rightarrow\infty} 0$.

This argument shows that all quantities $M, V, H, F$ converge to $0$.  

Finally, if $\dv{U}{t} = W$ with $W(t)\ge0$ and such that $U$ is bounded,  then $U$ is convergent. From equation \ref{eqm} we know that $W(t)\geq 0$, thus $U$ is an increasing function, which combined with the boundedness of $U$ 
guarantees that $U$ converges. \\

In conclusion,  the disease‐free equilibrium is globally asymptotically stable,
$$
   \lim_{t\to\infty}I(t)\;=\;\lim_{t\to\infty}E(t)\;=\;0,
   \quad
   \lim_{t\to\infty}S(t)\;=\;S^*<\tfrac{\gamma}{\beta},
 $$ 

and all other compartments driven by $I$ likewise vanish or converge.
\end{proof}

\end{document}